\documentclass[aoas]{imsart}

\RequirePackage{amsthm,amsmath,amsfonts,amssymb}
\RequirePackage[authoryear]{natbib}
\RequirePackage[colorlinks,citecolor=blue,urlcolor=blue]{hyperref}
\RequirePackage{graphicx}
\usepackage{longtable, stmaryrd, dsfont, wasysym}
\usepackage{algorithm}
\usepackage{algpseudocode}
\usepackage{hyperref}

\startlocaldefs
\theoremstyle{plain}

\newtheorem{theorem}{Theorem}[section]

\newtheorem{remark}[theorem]{Remark}
\theoremstyle{remark}

\endlocaldefs
\begin{document}
\begin{frontmatter}
\title{A Structured Estimator for large Covariance Matrices in the Presence of Pairwise and Spatial Covariates}
\runtitle{A Structured Estimator for large Covariance Matrices}

\begin{aug}
\author[A]{\fnms{Martin}~\snm{Metodiev}\ead[label=e1]{martin.metodiev@doctorant.uca.fr}},
\author[B]{\fnms{Marie}~\snm{Perrot-Dock\`{e}s}\ead[label=e2]{marie.perrot-dockees@u-paris.fr}},
\author[H]{\fnms{Sarah}~\snm{Ouadah}\ead[label=e3]{sarah.ouadah@sorbonne-universite.fr}},
\author[G]{\fnms{Bailey K.}~\snm{Fosdick}\ead[label=e4]{bailey.fosdick@cuanschutz.edu}},
\author[H]{\fnms{Stéphane}~\snm{Robin}\ead[label=e5]{stephane.robin@sorbonne-universite.fr}},
\author[A,F]{\fnms{Pierre}~\snm{Latouche}\ead[label=e6]{Pierre.LATOUCHE@uca.fr}} \and
\author[D]{\fnms{Adrian E.}~\snm{Raftery}\ead[label=e7]{raftery@uw.edu}}

\address[A]{Université Clermont Auvergne, Laboratoire de Mathématiques Blaise Pascal,\\\printead[presep={ }]{e1,e6}}
%\address[E]{Corresponding author\printead[presep={,\ }]{e1}}
\address[B]{Université Paris Cité, CNRS, MAP5, F-75006 Paris, France\printead[presep={,\ }]{e2}}
\address[H]{Sorbonne Université, Laboratoire de Probabilités, Statistique et Modélisation (LPSM)\printead[presep={,\ }]{e3,e5}}
\address[G]{GTI Energy\printead[presep={,\ }]{e4}}
\address[F]{Institut universitaire de France (IUF)\printead[presep={,\ }]{e6}}
\address[D]{University of Washington, Department of Statistics\printead[presep={,\ }]{e7}}
\runauthor{M. Metodiev et al.}
\end{aug}

\begin{abstract}
 We consider the problem of estimating a high-dimensional covariance matrix from a small number of observations when covariates on pairs of variables are available and the variables can have spatial structure. This is motivated by the problem arising in demography of estimating the covariance matrix of the total fertility rate (TFR) of $195$ different countries when only $11$ observations are available. We construct an estimator for high-dimensional covariance matrices by exploiting information about pairwise covariates, such as whether pairs of variables belong to the same cluster, or spatial structure of the variables, and interactions between the covariates. We reformulate the problem in terms of a mixed effects model. This requires the estimation of only a small number of parameters, which are easy to interpret and which can be selected using standard procedures. The estimator is consistent under general conditions, and asymptotically normal. It works if the mean and variance structure of the data is already specified or if some of the data are missing. We assess its performance under our model assumptions, as well as under model misspecification, using simulations. We find that it outperforms several popular alternatives. We apply it to the TFR dataset and draw some conclusions. 
\end{abstract}

\begin{keyword}
\kwd{large covariance matrix estimation}
\kwd{pairwise covariates}
\kwd{spatial effects}
\end{keyword}
\end{frontmatter}

\section{Introduction\label{sec: Introduction}}
%Define Yt and Ytj
    We consider the problem of estimating a large covariance matrix from a small number of data points when covariates on pairs of variables and their spatial structure are available. This is motivated by the problem, arising in demographic research, of estimating the covariance matrix of the total fertility rate for a large number of countries from data at a small number of time points.
    
    Our specific goal is to estimate the covariance matrix of a model for the total fertility rate (TFR) used by the United Nations (UN) for 195 different countries. The dataset is denoted $Y$ and is made up of measurements of the TFR at $T=11$ time points (observations) for $d=195$ countries (variables). Each time point is a five-year period.  The large dimension of the covariance matrix makes the use of a standard estimator, such as the sample covariance matrix, untenable, and necessitates the use of additional information. Fortunately, for the TFR dataset, we have been able to obtain information about pairwise, time-invariant, covariate structures. In particular, these include a spatial structure (e.g., countries being contiguous to each other), and structures determined by 
cluster membership  (e.g., countries belonging to the same region, countries having the same common colonizer). The framework we propose also allows us to include interactions when modeling a covariance matrix.

\paragraph*{\textbf{Covariance matrix estimation}}
As we shall see,  the performance of standard covariance matrix estimators is not excellent in the setting we consider, because they cannot handle pairwise and spatial covariates. These include shrinkage estimators, such as the Ledoit-Wolf estimator \citep{LeWo04-estimlargecovmat03}, which shrink covariance matrices towards sparse matrices or the identity matrix (for a review, see \citealp{Po13-highDimCovmatEstims_review, LeWo22-estimlargecovmat02}), as well as estimators such as the graphical lasso \citep{Fr_et_al08-graphical_lasso}, which assumes sparsity of the inverse of the covariance matrix. 

The most commonly used approaches which assume dependency structures within the model are approaches that use factor models \cite[for an overview of these approaches, see][]{Fa_et_al15-estimlargecovmat04}. There are indeed estimation procedures that use factor models and which can also allow for the encoding of spatial information, as in   \cite{ChAm03-multivariate_spatial_factoranalysis,WaWa03-generalized_spatial_factormodels,Ga_et_al08-spatial_dynamic_factoranalysis,Lo_et_al11-spatial_dynamic_factormodels,Th_et_al15-spatial_factoranalysis}. Clustered structures, on the other hand, are estimated via multilevel factor models \citep{NiMu92-multilevel_factor_analysis}. While we will show that the correlation structures of the different cluster effects present in the data can each be interpreted in the framework of multilevel factor models, none of the models available in the literature can combine several multilevel factor models with a spatial structure. Moreover, the number of samples we are facing is too small to use techniques such as parallel factor analysis \citep{HaLu94-parallel_factor_analysis}.

\paragraph*{\textbf{Interpretability}}

Interpretability of the model parameters has been stressed to be an important feature of covariance matrix parametrization by \citet{Po99-interpretable_unconstrained_parametrization,Po11-interpretable_unconstrained_parametrization02}, but while that work does provide solutions, they do not incorporate pair-specific covariates into the covariance matrix estimation procedure. Approaches that do include pairwise covariates into the model have involved evaluating linear combinations of known matrices on the scale of the covariance matrix \citep{An73-firstlinearcovarpaper}. Sums of covariance matrices also appear in linear mixed-effect models when adding crossed effects \citep{GaBu13-LMM_book}. 

However, sums of covariance matrices are, in practice, hard to interpret. To illustrate this problem, suppose that we set the covariance matrix of the data to the sum of the following two covariance matrices: \begin{align}\label{eq: covmat_combination}
    \Sigma:=\begin{pmatrix}18&0\\
    0&18
\end{pmatrix}+\begin{pmatrix}
    2& 0.8\\ 0.8 & 2
\end{pmatrix},\end{align} 
and that a two dimensional Gaussian model with covariance matrix $\Sigma$ is used to model the data. The first matrix assumes independence between the two variables considered, while the second matrix characterizes variables which are clearly correlated. Individually, the correlations of the second matrix tell us nothing about the correlation matrix of the data, since they might be overwhelmed by the entries of the first. Indeed, while the correlations of the two variables associated with the second matrix are relatively high (0.4), the  correlation matrix associated with $\Sigma$ is given by\begin{align}\label{eq: R_example}
    R=\begin{pmatrix}
        1&0.04\\0.04&1 \end{pmatrix},
\end{align} meaning that the correlation between the two variables is small in the resulting dataset.

Different scales, such as the matrix logarithm \citep{Ch_et_al96-covmatLoglink}, have been suggested, but they suffer from a similar problem.  \citet{BoJo16-MCGLM} generalized these approaches into a framework for non-normal multivariate data, called multivariate covariance generalized linear models (McGLM). A McGLM was used by \cite{BoJo16-MCGLM} to include spatio-temporal covariates in the covariance estimation. It is also possible to use \cite{Ca_et_al22-transitiveattributes} to incorporate information about variables belonging to a similar cluster. However, while the framework of McGLM is the closest to ours, it is not clear how to use it to get interpretable results, or how to combine spatial structures and clusters in the same model.

Some Bayesian approaches get interpretable parameters by separating the structure of the variance parameters from the structure of the correlation parameters. This separation strategy has been used successfully in Bayesian inference, in a variety of approaches \citep[see for instance ][]{Ba_et_al00-separationstrategy01,Le_et_al09-LKJprior, To_et_al11-covmatvisualization}. There are also some Bayesian approaches for covariance matrix estimation which use information about different clusters of variables, such as  \citet{Ka92-BayesianCorrEst_withGroups02,Ka93-BayesianCorrEst_withGroups,AgWe00-BayesianFactorModels02,Be_et_al03-BayesianFactorModels,Li_et_al04-BayesianCorrEsts}, but none of these combine this information with a spatial structure. 

At the core of the covariance matrix estimation strategy we propose is the idea of combining the separation strategy with the approach of \citet{An73-firstlinearcovarpaper}. This allows us to normalize the covariance matrices into correlation matrices, such that the parameters of the random effects are interpretable. Going back to Equations \eqref{eq: covmat_combination} and \eqref{eq: R_example}, we recommend instead separating the variance estimation and modeling $R$ directly as a convex combination of two correlation matrices. In our previous toy example, a possible choice could be:
\begin{align*}
    R=0.9\begin{pmatrix}
    1&0\\0&1
\end{pmatrix}+0.1\begin{pmatrix}
    1&0.4\\0.4&1
\end{pmatrix}.\end{align*} 
The importance of each matrix is given by its linear coefficient. Independently of the values of the other matrix, we will know that the correlation between the two variables is such that 10 percent of the correlation structure is explained by the second correlation matrix. 

In the TFR dataset that motivates our work, a variety of effects and their interactions are to be taken into account. We will show that the setting of the TFR dataset allows us to give useful estimates that work well, in the sense that useful properties (identifiability, consistency, asymptotic normality in the number of data points $T$ \textit{and} in the dimension $d$, available confidence regions) hold under some assumptions. We will also adjust our estimator in such a way that it gives consistent estimates, without having to assume that specific model assumptions hold.

\paragraph*{\textbf{Outline of the paper}}

The manuscript is organized as follows. In Section \ref{sec: Modelling Covariance Matrices}, we give an overview of the dataset and give the motivation for our model assumptions, followed by the general setting for the estimator we propose. In Section \ref{sec: Inference of Sigma}, we define our estimator and derive  its properties. We show how to estimate the parameters of our estimator (Section \ref{subsec: Model selection}) and how to adjust it in the case that the model assumptions do not hold (Section \ref{subsec: Model misspecification}). In Section \ref{sec: Numerical experiments} we assess its performance under a variety of different settings by simulation, and compare it to some popular alternative methods. Then, we use these techniques to derive estimates of the covariance matrix of the TFR dataset, whose correlations we compare in Section \ref{sec: Covariance estimates of the TFR dataset}. We close with a discussion of our work.  The code for this paper is made available via Github for scientific dissemination, see the following \href{https://github.com/M-crypto645/SCE}{link}.

\section{Modelling covariance matrices with known pairwise and spatial covariates\label{sec: Modelling Covariance Matrices}}

\subsection{The total fertility rate data\label{subsec: The total fertility rate covariance study}}

 The dataset we analyze consists of  the total fertilty rate (TFR) of 195 countries in successive $5$-year periods from 1950 to 2010. One country as well as one time period were removed for reasons on which we will elaborate further in this section and Section \ref{sec: Covariance estimates of the TFR dataset}. The dataset is denoted by $Y$ and is made up of $T=11$ time points (observations) and $d=195$ countries (variables). The element $Y_{t,j}$ of $Y$ is the TFR of country $j$ at time $t$, and $Y_t^{\intercal} := (Y_{t,1},\dots, Y_{t, d})\in\mathbb{R}^d$ denotes row $t$ of matrix $Y$. Note that the index $t$ is used to refer to different observations because of the specific nature of the dataset. There is previous work on this dataset, on which we build. We will first summarize this work. 

\cite{FoRa14-corlinearreg} found that  the model then used by the United Nations (UN) to predict the total fertility rate (TFR) of different countries yielded prediction intervals for forecasting regions consisting of multiple countries that were too narrow. They found that this was due to between-country dependencies that are not included in the Bayesian hierarchical model of \cite{Al_et_al11-tfr_projections}, which was used for these predictions. To tackle this, Fosdick and Raftery proposed estimating the covariance matrix of the joint distribution of the countries' TFRs.

They did this by modeling the covariance matrix using time-invariant pairwise information, namely whether two countries had the same common colonizer after 1945, whether two countries are contiguous, and whether two countries are in the same UN region\footnote{We identify regions by the UN subdivision of the sustainable development goal (SDG) regions into 21 geographic subregions. We identify the 10 different colonizers and their colonial relationships after 1945, as well as their neighborhood structure, via the database of the Centre d’\'{E}tudes Prospectives et d’Informations Internationales (CEPII) \citep{MaZi06-CEPIInotes}}.

A first model was built as
\begin{eqnarray}\label{Y_model}
Y_{t}&=& \mu_{t}+ \varepsilon^0_{t}\nonumber,\quad \varepsilon^0_{t} \sim \text{MVN}_d(0_d,\Sigma_t), 
\end{eqnarray}
with independent forecast errors $\varepsilon_t^0$, 
$\mu_{t}^{\intercal}:=(\mu_{t,1},\dots, \mu_{t,d})$, where $\mu_{t, j}$ denotes the expected TFR (conditional on $Y_{t-1}$) for country $j$ in period $t$ for $j\in\llbracket 1,d=195 \rrbracket$ and $\Sigma_t$ the covariance matrix (conditional on $Y_{t-1}$), $t\in\llbracket 1,T=11\rrbracket$. \citet{FoRa14-corlinearreg} also decomposed the covariance structure of the model into the standard deviation vector at time $t$, $\sigma_{t}\in\mathbb{R}^d_+$, as well as a correlation structure $R_t$. Conditional on $Y_{t-1}$, we have that:
\begin{equation*}
     \varepsilon^0_{t} \sim \text{MVN}_d(0_d,\text{diag}(\sigma_t) R_t \text{diag}(\sigma_t) ), 
\end{equation*} where $\text{diag}(\sigma_t)$ is a diagonal matrix with entries $\sigma_t$ and $\textrm{MVN}_d(m,S)$ denotes the multivariate normal distribution of dimension $d$ with mean $m$ and covariance matrix $S$.
 
With the method of \cite{Al_et_al11-tfr_projections} already providing accurate estimates of $\mu_t$ and $\sigma_t$, Fosdick and Raftery chose to model  $\varepsilon_{t}$, the standardized version of $\varepsilon^0_{t}$, such that 
\begin{equation} \label{eq: stand_errors}
    \varepsilon_{t} \sim\text{MVN}_d(0_d,R_t),
\end{equation}
and focused on the estimation of $R_t$. For all countries that are in phase II or III of the model of \cite{Al_et_al11-tfr_projections}, they set \begin{align}
\label{eq:cor_fora14}
    (R_t)_{i,j}= \begin{cases}
         1 \textrm{ if } i =j, \\
         \alpha^{(1)}_0 + \alpha^{(1)}_1 \mathrm{contig}_{i,j} +\alpha^{(1)}_2 \mathrm{comcol}_{i,j} + \alpha^{(1)}_3 \mathrm{region}_{i,j}\text{ if  }\; Y_{t,i}<\kappa\textrm{ and } Y_{t,j}< \kappa, \\
         \alpha^{(2)}_0 + \alpha^{(2)}_1 \mathrm{contig}_{i,j} +\alpha^{(2)}_2 \mathrm{comcol}_{i,j} + \alpha^{(2)}_3 \mathrm{region}_{i,j}\textrm{ otherwise,}
    \end{cases}
\end{align} where $\kappa$ is a threshold parameter, $\mathrm{contig}_{i,j} =1$ if countries $i$ and $j$ are contiguous, $\mathrm{comcol}_{i,j} =1$ if they had a common colonizer after 1945,  $\mathrm{region}_{i,j} = 1$ if they are in the same UN region, and $0$ otherwise.

The problem with this approach is that the value of $R_t$ that it yields is not necessarily positive definite for all values of $\alpha_0^{(k)},\dots,\alpha_3^{(k)}$. To address this, \cite{FoRa14-corlinearreg} mapped their estimate of $R_t$ onto the positive definite correlation matrix that was closest to it with regards to the Frobenius norm. This is quite simply done computationally, by taking the eigenvalue decomposition of the estimated matrix, setting negative eigenvalues to a small positive value, and reconstituting the matrix. However, this method comes with no statistical guarantees. We seek instead a statistically principled approach to estimating the covariance matrix. 

Here we propose a new approach that models the correlation matrix of the TFR directly, as a time-independent correlation matrix $R$, which is a linear combination of several correlation matrices. These correlation matrices are in turn entirely determined by known covariates, independent of the thresholding level $\kappa$, and in particular of the time point $t$, which is why we do not use the index $t$ for $R$ in our model. Therefore, $R$ will be positive definite if at least one of these correlation matrices is positive definite. 

For each time point $t$, we follow \citet{FoRa14-corlinearreg} in only modeling the TFR $Y_t$ of the countries that are in phase II or III of the model of \citet{Al_et_al11-tfr_projections}. This occurs only after the first time period, which is why one time period was removed to obtain $T=11$. 
Our methodology relies on the decomposition of the stanadardized errors $\varepsilon_t$ as a weighted sum of standardized, independent effects. Thus, for every time point $1\leq t\leq T$, we consider: \begin{eqnarray}\label{Y_model}
Y_{t}&=& \mu_{t}+ \text{diag}(\sigma_t)\varepsilon_{t}\nonumber,\quad \varepsilon_{t} \sim \text{MVN}_d(0_d,R). 
\end{eqnarray} 
We model the residuals as follows:
\begin{eqnarray}\label{epsilon_model}
\varepsilon_{t}&=&  A_{t} + B_{t} + C_{t} + D_{t} + E_{t},\textrm{  }A_t,B_t,C_t,D_t,E_t\in\mathbb{R}^d\textrm{ where}\label{eq:tfr_random_effect_model}\\A_t&\stackrel{\text{i.i.d}}\sim&{\rm MVN}_d(0_d,\alpha_AF_A),
\quad B_t\stackrel{\text{i.i.d}}\sim{\rm MVN}_d(0_d,\alpha_BF_B), 
\quad C_t\stackrel{\text{i.i.d}}\sim{\rm MVN}_d(0_d,\alpha_CF_C ), \nonumber\\ D_t&\stackrel{\text{i.i.d}}\sim&{\rm MVN}_d(0_d,\delta_D\Gamma(\beta_D)^{-1}), 
\quad E_t\stackrel{\text{i.i.d}}\sim{\rm MVN}_d(0_d,\alpha_E I_{d}).\nonumber
\end{eqnarray} 
Note that the covariance matrix of the standardized errors is equal to the correlation matrix of the data, conditional on the values at the previous time points, since by construction each $\varepsilon_{t,j}$ has variance 1.

The random vectors, $A_{t,j},B_{t,j},C_{t,j},D_{t,j}$ denote the random effects on country $j$ at time $t$, corresponding to the effect of having a common colonizer, belonging to the same region, the global effect, the contiguity effect, respectively. The random vector $E_{t}$ denotes i.i.d Gaussian noise for all countries at time $t$ (its correlation matrix is the identity matrix of dimension $d$, $I_d$). Its presence ensures that the correlation matrix of $\varepsilon_t$ is positive definite, since at least one of the correlation matrices of the standardized effects is positive definite. The random effects are standardized, in the sense that the matrices $F_A,F_B, F_C,\Gamma(\beta_D)^{-1}$ are correlation matrices, i.e.,\ positive semidefinite matrices with diagonal entries equal to 1. They are also weighted by positive constants $\alpha_A,\dots,\alpha_E,\delta_D$, which must sum to one. This ensures that the impact of each effect is measurable via its linear coefficient: for every pair of countries $(i,j)$, its correlation is an average of the individual entries $(F_A)_{i,j},(F_B)_{i,j},(F_C)_{i,j},\Gamma(\beta_D)^{-1}_{i,j}$ weighted by their respective weights. 

 We use two separate approaches for modeling the correlation matrices of the random effects. Matrices with the capital letter $F$ naturally impose a block structure on the model, in the sense that they partition the countries based on which cluster (i.e.,\ region, colonizer) they belong to. Let $f_A\in\{0,1\}^{d\times 10},f_B\in\{0,1\}^{d\times 21}$ denote the covariates corresponding to the regional and common colonizer clusters, meaning that $(f_B)_{j,r}=1$ if country $j$ belongs to region $r$ and $0$ otherwise (analogously for $f_A$). We can easily transform these covariates into correlation matrices by setting  $F_A=f_A f_A^\intercal,F_B=f_Bf_B^\intercal$, meaning that $(F_A)_{i,j},(F_B)_{i,j}$ are equal to 1 if countries $i$ and $j$ have the same common colonizer or belong to the same UN region, respectively. The global effect applies equally to every country pair, such that $F_C=1_d1_d^\intercal$, where $(1_d)_j=1$ for all $j$, meaning that $(F_C)_{i,j}$ is always 1. Note that the way these matrices are modeled corresponds to the way in which correlation matrices are modeled in multilevel factor models, only that in our case the factor loadings and group means are known. They can also be represented via the design matrix of a linear mixed-effects model. We elaborate on these points in Appendix D.

The contiguity effect was modeled using a conditional autoregressive (CAR) model. There are many different CAR models, each of which is usually specified by a small number of parameters \citep{Wa04-CARexplanations,Ky_et_al10-DCARmodel, Ma11-CARvariations, Ta_et_al13-DCARspatiotemporal, Ve_et_al18-CARmodeltheory}. We parametrize the CAR model via only one autocorrelation parameter, $\beta_D$, in a Gaussian Markov random field (GMRF). We follow \cite{BeKo95-CAR_GMRF_01} and \cite{Be_et_al91-CAR_GMRF_02} in setting \begin{align}
\label{eq:gamma}
    \Gamma(\beta_D)=Q_{\beta_D} M_2(I_d-\beta_D M_1)Q_{\beta_D},\; (M_2)_{i,j}=\begin{cases}
        \sum_{e=1}^d M_{i,e}&i=j\\0&i\neq j
    \end{cases},\; (M_1)_{i,j}=\frac{M_{i,j}}{(M_2)_{i,j}},
\end{align} with the parameter $\beta_D\in(0,1)$, where $M$ is the adjacency matrix induced by the neighborhood structure of the underlying geography. The expression for $Q_{\beta_D}$ in (\ref{eq:gamma}) is different from the most commonly used form. This is because here we are modeling the correlation matrix rather than the covariance matrix. The matrix $Q_{\beta_D}$ is a non-negative diagonal matrix chosen such that all diagonal entries of $\Gamma(\beta_D)^{-1}$ are equal to 1. In the model we propose, a simple analytic expression of $Q_{\beta_D}$ exists, as illustrated in Appendix A.

It can happen that components of the CAR model are unconnected, and that there are countries that have no neighbors at all (e.g., countries that consist of islands). The latter case is of particular importance to us since the standard CAR model is not defined for unconnected nodes. However, the literature on how to deal with this case is sparse. We follow \cite{Fr_et_al18-disconnected_plus_singletons} in assuming that the correlation of the spatial effect is $\Gamma(\beta_D)^{-1}_{j,j}=1$ for every country $j$ and $\Gamma(\beta_D)^{-1}_{i,j}=0$ if country $i$ or country $j$ is set on an island.

Note that the assumptions made in Equation \eqref{epsilon_model} are general: we assume only that there are some independent, standardized effects through which we can express the impact of our covariates on the correlation matrix of the data.

\subsection{Correlation structure\label{subsec: Correlation structure}}

%Let us consider a multidimensional Gaussian vector $Y\sim\mathcal{N}(\mu,\text{diag}(\sigma) {\color{blue}R(\alpha,\beta,\delta)} \text{diag}(\sigma))$.
We now develop a more general framework for correlation estimation.
Suppose that there are $K$ known correlation matrices $F_1,\dots,F_K$ defined with no parameters (these may correspond, for example, to correlation matrices derived from clusters, such as regions and colonizers), one known positive definite correlation matrix $F_0$ (this will usually correspond to i.i.d Gaussian noise) and one known correlation matrix defined via a number of parameters $\beta_1,\dots,\beta_G$ (this will usually correspond to the correlation matrix of a spatial effect). Let $Y_1,\dots,Y_T$ be a Markov process of Gaussian vectors with mean vectors $\mu_t=\text{E}[Y_t|Y_{t-1}]$, variance vectors $\sigma_t^2=\text{Var}[Y_t|Y_{t-1}]$, and a correlation matrix $R=\text{Cor}(Y_t|Y_{t-1})$, which does not depend on the time point $t$. We model the correlation matrix $R$ as follows: 
\begin{equation} \label{model}
R(\alpha,\beta,\delta) = \Phi(\alpha) + \delta\cdot \Gamma(\beta)^{-1}, \qquad
\text{where} \quad
\Phi(\alpha) = \sum_{k=0}^K \alpha_k F_k, \quad
%\quad \alpha_k,\beta_\ell>0,
\end{equation}
with the constraints \begin{equation}
    \alpha_k,\delta>0,\quad\delta+\sum^K_{k=0}\alpha_k=1.
\label{eq:general_simplex_restriction}\end{equation}

Since the parameters must sum to 1, we restrict our estimation procedure to only estimating $(\alpha,\beta,\delta):=(\alpha_1,\dots,\alpha_K,\beta_1,\dots,\beta_G,\delta)$, because one can easily solve for $\alpha_0=1-\sum_{k=1}^K \alpha_k-\delta$.

Here we focus on estimating the correlation matrix $R$ by modeling it as a sum of correlation matrices. One could also estimate the covariance matrix $\Sigma$ by modeling it as a sum of covariance matrices.  The same kind of methodology can still be applied in this case, with some minor changes. 

Few assumptions about $\Gamma(\beta)$ are needed, other than assumptions on its differentiability. The order of differentiability required for our algorithm to work well will be discussed in the next section. Ideally, $\Gamma(\beta)^{-1}$ should be twice differentiable with bounded first derivatives and continuous second derivatives.

Model \eqref{model} is a generalization of model  \eqref{eq:tfr_random_effect_model}: due to the assumption of independence we know that the correlation matrix of the standardized errors is a linear combination of the correlation matrices of the random effects. Also, the diagonal entries of a correlation matrix are equal to 1. The restriction \eqref{eq:general_simplex_restriction} ensures that this is the case for the model we propose. 
 
The structure is defined on the scale of the correlation matrix. No assumptions are made on the structure of the variance vectors $\sigma_t^2$ or the mean vectors $\mu_t$. They may even depend on the data, but only on the data of their predecessor, $Y_{t-1}$. Thus the data follow the following distribution: \begin{equation}
\label{eq:log_lik}
   (Y_1,\dots, Y_T) \sim \prod^T_{t=1}\text{MVN}_d(\mu_t, \Sigma(\sigma_t,\alpha,\beta,\delta)),\;\; \Sigma(\sigma_t,\alpha,\beta,\delta)= \text{diag}(\sigma_t)R(\alpha,\beta,\delta)\text{diag}(\sigma_t),
\end{equation}
where $R(\alpha,\beta,\delta)$ is defined in Equation \eqref{model}. 

Note that the description of the data was chosen to be as general as possible, allowing dependencies between the data points, as well as mean- and variance structures that vary with time. This allows us to add a correlation structure to models which are potentially quite complicated, but already have well-known mean- and variance estimators available. 
    
This is the case for our model, in which the countries marginally follow a random walk model with nonlinear drift, and for which the marginal mean- and variance parameters have already been estimated efficiently in \citet{Al_et_al11-tfr_projections}. This is also why we are primarily interested in the properties of our estimator on the scale of the correlation matrix. 
    
However, one could also use our model in the classical case, where all data points are i.i.d and use standard estimators for the mean- and variance parameters, in which case consistency on the scale of the correlation matrix carries over to consistency on the scale of the covariance matrix. We implement an algorithm for this case and test it in Section \ref{sec: Numerical experiments}.

\section{Inference about the covariance matrix\label{sec: Inference of Sigma}}
\subsection{\label{subsec: Maximum Likelihood estimation}Maximum likelihood estimation}

The likelihood associated with the proposed model is 
\begin{align}L(Y;\mu,\sigma,\alpha,\beta,\delta)=\prod^{T}_{t=1}\text{MVN}_d(Y_t;{\mu_t},\Sigma(\sigma_t,\alpha,\beta,\delta)),
\label{eq:likelihood}\end{align} 
where $\mu=(\mu_t)_t,\sigma=(\sigma_t)_t$.
In our context, accurate estimates of $\mu$ and $\sigma$ are already available and denoted by $\hat{\mu}$ and $\hat{\sigma}$ respectively. We denote the maximum likelihood estimator of $\Sigma(\sigma_t, \alpha,\beta,\delta)$ by: \begin{align*}
\hat{\Sigma}^{\text{SCE}}_t &:=\Sigma(\hat{\sigma}_t,\hat{\alpha}_{\text{SCE}},\hat{\beta}_{\text{SCE}},\hat{\delta}_{\text{SCE}}),
\end{align*}
where $(\hat{\alpha}_{\text{SCE}},\hat{\beta}_{\text{SCE}},\hat{\delta}_{\text{SCE}} )$ is the output of an optimization algorithm that solves\begin{align*}
(\hat{\alpha}_{\text{SCE}},\hat{\beta}_{\text{SCE}},\hat{\delta}_{\text{SCE}} )& = \textrm{argmax}_{(\alpha,\beta,\delta)}L(Y;\hat{\mu},\hat{\sigma},\alpha,\beta,\delta).
\end{align*} 
We call this  the "structured covariance estimator" (SCE), because it is meant to capture known pairwise dependency structures in the data.
Note that if one does not already have accurate estimates of $\mu$ and $\sigma$ it is also possible to optimize the likelihood in Equation~\eqref{eq:likelihood} jointly over $(\mu,\sigma,\alpha,\beta,\delta)$. 

\paragraph*{\textbf{Identifiability}}
Before computing the SCE one might check that the model is identifiable.
This is relatively easy to do if the dimension of $\beta$ is small (note that in our case $\beta$ is one-dimensional). This is because for fixed $\beta$, identifiability can be checked by solving a linear program.

\begin{theorem}\label{thm-identifiability}

The model is identifiable if the linear program \begin{equation*}
\begin{array}{ll@{}ll}
\mathrm{max}_{\alpha,\alpha',\delta,\delta'}  & \delta+\delta'\\
\mathrm{such}\; \mathrm{that } &\displaystyle\sum\limits_{k=0}^{K}(\alpha_k-\alpha_k')F_k+\delta\Gamma(\beta)^{-1}    = \delta'\Gamma(\beta')^{-1}\\&\displaystyle\sum\limits_{k=0}^{K}\alpha_k+\delta=\displaystyle\sum\limits_{k=0}^{K}\alpha_k'+\delta'=1,\quad\alpha_0,\dots,\alpha_K,\alpha_0',\dots,\alpha_K',\delta,\delta'\geq0
\end{array}
\end{equation*} has output 0 and the matrices $F_0,\dots,F_K,\Gamma(\beta)^{-1}$ are linearly independent for every $\beta\neq\beta'$.
\end{theorem}

The proofs of this and any other theorem in this article can be found in Appendix B. The theorem can be used to verify identifiability, since its conditions can be checked with a grid-search over $\beta,\beta'$. Let us note that this check will return parameter values for which the model is not identifiable, if it finds any.

\paragraph*{\textbf{Initialization}}
In order to inialize the procedure, we consider \begin{align}
\label{eq:init}
(\alpha^{(0)},\beta^{(0)},\delta^{(0)}):&=\text{argmin}_{(\alpha,\beta,\delta)}\|R(\alpha,\beta,\delta)-\hat{R}\|_F,\end{align} where  $R(\alpha,\beta,\delta)$ is defined in Equation \eqref{model}, $\|.\|_F$ denotes the Frobenius norm and $\hat{R}$ is the Pearson correlation matrix of the estimated standardized errors, $\hat{\varepsilon}_t=\text{diag}(\hat{\sigma}_t)^{-1}(Y_t-\hat{\mu}_t)$. This particular initialization is useful because $R(\alpha^{(0)},\beta^{(0)},\delta^{(0)})$ itself is a consistent estimator as long as $\hat{R}$ is consistent.

For fixed $\beta$, the initialization is an optimization problem that can be solved in polynomial time, as long as the conditions of Theorem \ref{thm-identifiability} hold, using a convex quadratic optimization method from \citet{GoId83-quadratic_optimization02, GoId06-quadratic_optimization01}, implemented in the R-package quadprog \citep{BeWe19-quadprog}. Thus, we can  solve this problem via a grid-search over $\beta$.

Note that the optimal solution of Equation \eqref{eq:init} may lie on the edge of the parameter space. This is a problem since parameter values such as $\alpha_1=1$ are not allowed. In this case, additional constraints are added to ensure that the optimal solution is feasible. The complete procedure used is defined in Appendix C.

\paragraph*{\textbf{Gradient descent}}
It is equivalent, and computationally more convenient, to maximize a specific transformation of the likelihood, denoted by $l(\alpha,\beta,\delta)$, which is defined in Appendix A. We do this by using a quasi-Newton algorithm (BFGS) \citep{FlRe64-optimDefinitionBFGS,Br70-BFGS02, Go70-BFGS03,Sh70-BFGS04}. This algorithm requires derivatives of $l$, which are easy to obtain in our case since there is a simple expression for the spatial effect matrix $\Gamma(\beta_D)^{-1}$. A list of these derivatives is given in Appendix A.

\subsection{\label{subsec: MLE Properties}MLE Properties}

In Theorem~\ref{thm-consistent_in_T}  below, we prove several desirable properties of this estimator, such as consistency and asymptotic normality in the number of time points $T$. Interestingly, it turns out that under mild conditions, we also have consistency and asymptotic normality in the dimension of the data, $d$, even if there is just one time point. This result is given in Theorem~\ref{thm-consistent_in_n}. This is important because in the case of the TFR dataset the data are observed for at most $d=195$ countries and $T=11$ time points.

\begin{theorem}\label{thm-consistent_in_T}
    Let $(\alpha^{\star},\beta^{\star},\delta^{\star})$ be the true model parameter and $\Sigma^{\star}_t$ denote the true covariance matrix, $R^\star$ the true correlation matrix. Suppose that 
\begin{itemize}
    \item [(A1)] the model given by Equation \eqref{model} is identifiable,
 \item [(B1)] $\Gamma(\beta)^{-1}$ is a uniformly continuous function in $\beta$ with open,
convex and bounded domain,
 \item [(C1)] the squared estimated standardized errors $S_T=\frac{1}{T}\sum^T_{t=1}\hat{\varepsilon}_t\hat{\varepsilon}_t^\intercal$ almost surely converge to $R^\star$.
        \end{itemize}
        Then the normalized SCE, $\hat{R}_{\textup{SCE}}=R(\hat{\alpha}_{\textup{SCE}},\hat{\beta}_{\textup{SCE}},\hat{\delta}_{\textup{SCE}})$, is strongly consistent in the number of time points $T$.
        
If assumptions (A1)-(C1) hold and
                \begin{itemize}
                \item[(A2)]
$(\hat{\alpha}_\textup{SCE},\hat{\beta}_\textup{SCE},\hat{\delta}_\textup{SCE})$ is a stationary point of $l$,
            \item [(B2)]  $\Gamma(\beta)$ is at least twice differentiable in $\beta$ with continuous partial derivatives,
            \item[(C2)]   $Z_T:=\sqrt{T}(S_T-R^\star)$ converges to a multivariate normal distribution with mean $0_d0_d^T$ and covariances given by the correlations of $Y_1 Y_1^\intercal$,
        \end{itemize} then the vector $\sqrt{T}((\hat{\alpha}_{\textup{SCE}},\hat{\beta}_{\textup{SCE}},\hat{\delta}_{\textup{SCE}})-(\alpha^\star,\beta^\star,\delta^\star))$ is asymptotically normally distributed with means 0 and covariance matrix equal to the inverse of the Fisher information matrix \begin{align*}
            I(\alpha,\beta,\delta)_{i,j}=\frac{1}{2}\textup{tr}\left(R(\alpha,\beta,\delta)^{-1}\frac{\partial R}{\partial(\alpha,\beta,\delta)_i}(\alpha,\beta,\delta)R(\alpha,\beta,\delta)^{-1}\frac{\partial R}{\partial(\alpha,\beta,\delta)_j}(\alpha,\beta,\delta)\right)
        \end{align*} evaluated at the true parameter, in the number of time points $T$.

\end{theorem}

\begin{remark}
    The correlation matrix of the CAR-model, which we use in the experiments section (Section \ref{sec: Numerical experiments}) of this article, fulfills Conditions (B1) and (B2). We can check for identifiability (Condition A1), using Theorem \ref{thm-identifiability}. As time passes, the countries eventually enter phase III of the model of \citet{Al_et_al11-tfr_projections}. Thus, the majority of the mean and variance estimators used by \citet{FoRa14-corlinearreg} are, for $T$ sufficiently large, approximating MLEs of the autocorrelation and variance parameter of an AR(1) model, since only phase III of their model is relevant for consistency in $T$.  These types of estimators are strongly consistent and asymptotically normal for a very general class of models that use time series \citep{Ha73-asconvergence_timeseries_params} and the strong law of large numbers still holds for weakly correlated observations \citep{Ly88-strongLLN_weaklycorrelated}, so there is an argument to be made that Conditions (C1) and (C2) hold as well. Additionally, asymptotic confidence regions can be obtained by calculating the inverse of the Fisher information matrix. We can observe the properties of Theorem \ref{thm-consistent_in_T} in the experiments section.
\end{remark}

In our case, we are going to show in Section \ref{subsec: Performance in settings with different covariance structures} of this article that the number of time points necessary for convergence is quite small, as long as the dimension $d$ of the data is sufficiently large. In fact, thanks to the following theorem, we have convergence, even if $T=1$. 

\begin{theorem}\label{thm-consistent_in_n}
    Suppose that the estimated standardized errors follow an i.i.d normal distribution with mean $0_d$, unit variances and correlation matrix $R$, and that a global maximum of the likelihood is given by  $(\hat{\alpha}_\textup{SCE},\hat{\beta}_\textup{SCE},\hat{\delta}_\textup{SCE})$. Suppose also that
    \begin{itemize}
        \item[(A)] $\Gamma(\beta)$ is two times differentiable with respect to $\beta$ with continuous partial derivatives and open domain,
    
        \item[(B)] the model given by Equation \eqref{model} is ``identifiable in its limit" in $d$, in the sense that the limit of the correlation matrix of the Fisher information matrix exists and is non-singular, 

          \item[(C)] $F_0=I_d$, the sum of each row of the element-wise absolute values \\$|F_1|,\dots,|F_K|,|\Gamma(\beta)^{-1}|,|\frac{\partial}{\partial\beta_g}|\Gamma(\beta)^{-1}|,\frac{\partial^2}{\partial\beta_g\partial\beta_w}|\Gamma(\beta)^{-1}|$ is bounded in $d$ (i.e., it is $O(1)$) and for a proportion of at least $p\in(0,1)$ of the rows there exists a $\tau>0$ such that each sum of squares of the non-diagonal row-elements has a lower-bound of\ $\tau$.
    \end{itemize}  
    Then the estimators $(\hat{\alpha}_{\textup{SCE}},\hat{\beta}_{\textup{SCE}},\hat{\delta}_{\textup{SCE}})$ are consistent and $\sqrt{I(\hat{\alpha}_{\textup{SCE}},\hat{\beta}_{\textup{SCE}},\hat{\delta}_{\textup{SCE}})}((\alpha^\star,\beta^\star,\delta^\star)-(\hat{\alpha}_{\textup{SCE}},\hat{\beta}_{\textup{SCE}},\hat{\delta}_{\textup{SCE}}))$ converges in distribution in $d$ to a standard normal random variable. 
\end{theorem}

\begin{remark}
The theorem holds for any value of $T$, the number of time points, even for $T=1$. Conditions (A) and (C) are fulfilled by the CAR model, as long as the size of the largest connected component of the underlying spatial structure is bounded in $d$ and there are fewer island-countries than non-island-countries. In the case of the TFR dataset, Conditions (B)-(C) boil down to conditions on the data having multiple distinct clusters, each of which is limited in its size, but has more than one component. This is mostly the case, as there are several regions, colonizers and (approximately) separate connected components (i.e., continents). It is not fulfilled by the global effect, since its cluster is of size $d$. However, we found that the global effect only made a small impact in practice and thus should not affect performance too much. We analyze the convergence of our model in the number of countries, $d$, in Section \ref{subsec: Performance with varying dimensions}.  
\end{remark}

\subsection{\label{subsec: Model selection}Model selection} 

All the available matrices may not be relevant for modeling the covariance of the data.
If this is the case, only a subset of the matrices $\{\Gamma(\beta)^{-1},F_1,\dots,F_K\}$ should be used to construct the model. 

For an index set $J\subseteq\{-1,0,1,\dots,K\}$, we define
\begin{equation*}
R_J(\alpha,\beta,\delta) = \Phi_J(\alpha) + \delta\cdot \Gamma(\beta)^{-1}\cdot \mathds{1}_{-1 \in J}, \quad
\text{where} \quad
\Phi_J(\alpha) = \sum_{k\in J\backslash\{-1\}} \alpha_k F_k.
\end{equation*}
Then, we conduct model selection via the Bayesian information criterion (BIC). For any given index set $J$, we define the BIC as
\begin{align*}
    -2\log(L_J(Y;\hat{\mu},\hat{\sigma},\hat{\alpha}_{\textup{SCE}},\hat{\beta}_{\textup{SCE}},\hat{\delta}_{\textup{SCE}}))+(|J|+G\cdot \mathds{1}_{-1 \in J})\log(T),
\end{align*} 
with $L_J$  defined like $L$ in Equation~\eqref{eq:likelihood} and $R$ replaced by $R_J$ in the definition of $\Sigma$.
 The asymptotic normality of the model parameters, guaranteed by Theorems \ref{thm-consistent_in_T} and \ref{thm-consistent_in_n}, ensures that, under mild conditions, the posterior distribution is approximately normal, which justifies the use of the BIC \citep{Ra95-BIC_approx}.

\subsection{\label{subsec: Model misspecification}Model misspecification}

In some cases, only parts of the correlation coefficients are explained by known covariates. In such cases Equation~\eqref{model} does not hold and the SCE is not  consistent. However, the available covariates may still be useful to improve the efficiency of the covariance matrix estimation.

To capture both the part of the correlation that is due to the known covariates and the part that is not, we combine $\hat{R}_{\text{SCE}}$ with another correlation estimator  that does not make any assumption on the covariance structure. 

This is the case for the Pearson correlation estimator. In our specific setting, we have access to $\hat{\mu}$ and $\hat{\sigma}$, so we propose using a Pearson-type estimator:\begin{align*}
     (\hat{R}_{\text{Pearson}})_{i,j}=\begin{cases}\frac{1}{T-1}\sum^T_{t=1}\frac{(Y_{t,i}-\hat{\mu}_{t,i})(Y_{t,j}-\hat{\mu}_{t,j})}{\hat{\sigma}_{t,i}\hat{\sigma}_{t,j}}&i\neq j,\\1&i=j.\end{cases}
 \end{align*}

Note that if $\hat{\mu}$ and $\hat{\sigma}$ correspond to the sample mean and variance, $\hat{R}_{\text{Pearson}}$ is equal to the Pearson correlation matrix. If not, it may not be positive definite, in which case we map it to the positive definite correlation matrix that is closest in Frobenius norm, using the algorithm of \citet{ChHi98-nearPDalgorithm}, implemented in  the ``Matrix'' R package \citep{nearPD_package}. 

The convex combination of this estimator and $\hat{R}_{\text{SCE}}$ gives the estimator
\begin{align}
  \hat{\Sigma}^{\text{WSCE}}_t=\text{diag}(\hat{\sigma}_t)\hat{R}_{\text{WSCE}}\text{diag}(\hat{\sigma}_t),\quad \hat{R}_{\text{WSCE}}=(1-\hat{\lambda}_{\text{WSCE}})\hat{R}_{\text{SCE}}+\hat{\lambda}_{\text{WSCE}}\hat{R}_{\text{Pearson}}.
\label{eq:WSCE}\end{align}

$\hat{\Sigma}^{\text{WSCE}}_t$ is  consistent as long as $\hat{\lambda}_{\text{WSCE}}\in[0,1]$ approaches a given positive constant. \citet{LeWo03-CovMatCombination} make a very similar argument and give optimality conditions for the optimal mixing constant between two estimators, one of which may not be consistent. The same results hold in our setting, under the following conditions. \begin{theorem}\label{thm-modelmisspec}
    Suppose that all conditions of Theorem \ref{thm-consistent_in_T} hold and \begin{itemize}
        \item[(A3)] the model is misspecified, meaning that $R^\star\neq R(\alpha,\beta,\delta)$ for all $(\alpha,\beta,\delta)$, where $R^\star$ is the true, non-degenerate correlation matrix, 

        \item[(B3)] $(\hat{\alpha}_{\textup{SCE}},\hat{\beta}_{\textup{SCE}},\hat{\delta}_{\textup{SCE}})$ converges almost surely to its limit, $(\alpha^\star,\beta^\star,\delta^\star)$, and the $L^2$ limit of $\sqrt{T}((\hat{\alpha}_{\textup{SCE}},\hat{\beta}_{\textup{SCE}},\hat{\delta}_{\textup{SCE}})-(\alpha^\star,\beta^\star,\delta^\star))$ exists,

        \item[(C3)]  $\beta^\star$ is in the domain of\; $\Gamma$,

        \item[(D3)] $\sqrt{T}(S_T-R^\star)$ converges in $L^2$ to a multivariate normal-distributed random variable with mean $0_d0_d^\intercal$ and covariances given by the correlations of $Y_1 Y_1^\intercal$.
    \end{itemize} Then, the constant $\hat{\lambda}_{\textup{WSCE}}^\ast$ in Equation \eqref{eq:WSCE}, which minimizes the expected squared error of $\hat{R}_{\textup{WSCE}}$ in the Frobenius norm, is given by\begin{align*}
        1-\hat{\lambda}_{\textup{WSCE}}^*=\frac{1}{T}\cdot\frac{\pi-\rho}{\gamma}+O(\frac{1}{T^2})=O(\frac{1}{T}),
    \end{align*} where \begin{align*}
    \begin{pmatrix}\pi\\\rho\\\gamma\end{pmatrix}=  \sum^d_{i,j}\begin{pmatrix}
        \textup{\textbf{Var}}[\sqrt{T}(\hat{R}_{\textup{Pearson}})_{i,j}]\\\textup{\textbf{Cov}}[\sqrt{T}(\hat{R}_{\textup{SCE}})_{i,j},\sqrt{T}(\hat{R}_{\textup{Pearson}})_{i,j}]\\\left(\textup{\textbf{E}}[(\hat{R}_{\textup{SCE}})_{i,j}]-R^\star_{i,j}\right)^2
    \end{pmatrix}.
    \end{align*}
\end{theorem}

$\hat{\lambda}_{\text{WSCE}}^\ast$ can be thought of as a shrinkage constant. If the asymptotic expected error of the SCE, $\gamma$, is small, and if the asymptotic variance of the Pearson-type correlation matrix, $\pi$, is large, the WSCE is shrunk towards the SCE. Otherwise it is shrunk towards the Pearson-type correlation matrix. $\hat{\lambda}_{\text{WSCE}}^*$ approaches 1 as the sample size $T$ increases, which in turn ensures consistency of the WSCE due to the consistency of the Pearson-type correlation matrix.

Empirical estimates for the asymptotic expectations and variances used to define $\pi$ and $\gamma$ are readily available via standard estimation procedures (we give our choice in Appendix C). However, $\rho$ is defined via the covariances of $\hat{R}_{\text{SCE}}$ and the Pearson-type correlation matrix. These may be estimated via a bootstrap approach, but this would require a repeated calculation of the SCE over a large simulated sample, which is computationally expensive. Alternatively, one could approximate an upper bound of $\rho$, $\rho_U$, which is given by the Cauchy-Schwartz inequality:
\begin{align*}
    \rho_U=\sum^d_{i,j}\sqrt{\textbf{Var}[\sqrt{T}(\hat{R}_{\text{SCE}})_{i,j}]}\sqrt{\textbf{Var}[\sqrt{T}(\hat{R}_{\text{Pearson}})_{i,j}]}.
\end{align*} 
This upper bound approaches $\rho$ as the correlation between the two estimators increases. It is also easy to estimate, since estimators of the asymptotic variances of the Pearson correlation are well known, and the Fisher information can still be used to compute the variance of the SCE under model misspecification. One can thus make a choice between setting\begin{align*}
    \hat{\lambda}_{\text{WSCE}}^{\text{U}}=1-\frac{1}{T}\cdot\frac{\hat{\pi}-\hat{\rho}_U}{\hat{\gamma}},\quad\text{ or }\quad\hat{\lambda}_{\text{WSCE}}=1-\frac{1}{T}\cdot\frac{\hat{\pi}-\hat{\rho}}{\hat{\gamma}},
\end{align*} 
where the former expression can be computed efficiently, while the latter is more precise, but also more expensive to compute. In either case, $\hat{R}_{\text{WSCE}}$ is a consistent estimator of the correlation matrix of the data, under no assumptions on its correlation structure. 
 
     Note that it is possible that $\hat{\lambda}_{\text{WSCE}}^{\text{U}}$ and/or $\hat{\lambda}_{\text{WSCE}}$ lie outside of $[0,1]$. In this case, they are rounded to $0$ or $1$. $\hat{\lambda}_{\text{WSCE}}^{\text{U}}$ is also an upper bound on the optimal shrinkage constant, meaning that the WSCE can be shrunk too highly towards the Pearson-type correlation matrix. This can be avoided by computing $\hat{\lambda}_{\text{WSCE}}$ by using a bootstrap algorithm. We used this approach when the dataset contained missing values or when the mean and variance estimators were inprecise. However, we found that the upper bound that we suggested is quite accurate when the mean and variance vectors of the model are considered known and there are no missing values present, as illustrated in Section \ref{subsec: Performance under model misspecification} below. 

\section{Numerical experiments\label{sec: Numerical experiments}}

This section presents numerical experiments to emphasize the advantages and limits of the  proposed estimators compared to several state-of-the-art methods.
First, we compare their performances within the model assumptions, with varying covariate structures and sizes for the datasets. Then, we test the robustness of the model under missing values and model misspecification.

\subsection{Simulation settings}

\paragraph*{\textbf{Sample distribution}}

For each scenario, we simulated 40 independent datasets. Each dataset contains $T =11$ samples  drawn independently, such that \begin{equation}
\label{eq :simu}
    Y_{t} \sim \text{MVN}_d(0_d,R) \;  \text{for } 1 \leq t \leq T, 
\end{equation}
 with $R$ denoting a correlation structure from Equation \eqref{epsilon_model}. Details of all simulation settings can be found in Appendix E.
 
 Depending on the scenario, we considered  one of, or a combination of, the following two settings:
 \begin{itemize}
      \item \textbf{Fully simulated setting (FSS):} For each country, the membership vector indicating its region (resp. colonizer) is drawn from a multinomial distribution. The adjacency matrix of the spatial structure is simulated from an Erd\H{o}s-Rényi random graph model \citep{ErRe1959randomgraphs} with connection probability $\log(d)/d$. 
    \item \textbf{Using the TFR data (TFR):} $F_A,F_B, F_C$ and the adjacency matrix that defines the function $\Gamma(\beta_D)$ are chosen from the real data. %and coefficients from~\cite{FoRa14-corlinearreg}.
 \end{itemize}
 %with $d$ depending on the scenario.
\begin{remark}
 In the TFR dataset, the data points are not i.i.d. However, the standardized errors, which we use to estimate the covariance matrix, are approximately i.i.d, so this general setting is appropriate. Moreover, simulating the data as i.i.d allowed us to compare our estimators to 
 many standard estimators which can only work in an i.i.d setting. 
\end{remark}
 
\paragraph*{\textbf{External information}}

The SCE and the WSCE depend on estimates of $\mu$ and $\sigma$. These can be computed beforehand using external information. Thus, we distinguish two cases: 
\begin{itemize}
    \item  \textbf{Known}, where the true $\mu$ and $\sigma$ were used to calculate the SCE and the WSCE.
    \item  \textbf{Unknown}, where  $\mu$ and $\sigma$ were estimated by  empirical estimators.
\end{itemize}

\begin{remark}
 \cite{Al_et_al11-tfr_projections} provide accurate estimates of $\mu$ and $\sigma$ for the TFR dataset. Within their model, $\mu$ and $\sigma$ were estimated with only $15+3=18$ parameters (hyper-parameters of the Bayesian hierarchical model of phase II \citep{Al_et_al11-online_resource_01} plus parameters of the AR(1)-model of phase III) and not $2\cdot d=390$, which are needed for the empirical estimators in the unknown setting. 
Thus, the performance of the SCE (resp. WSCE) is likely to be in between the known and unknown case.
\end{remark}

\paragraph*{\textbf{Comparison estimators}}
For comparison, we considered the initial value estimator (IVE): 
\begin{equation} \label{eq:IVE}
    \hat{\Sigma}_t^{\text{IVE}} =\Sigma(\hat{\sigma}_t,\alpha^{(0)},\beta^{(0)},\delta^{(0)}),
\end{equation}
where $(\alpha^{(0)},\beta^{(0)},\delta^{(0)})$ are determined from the initialization step. 

We also computed Pearson's correlation matrix estimator, as well as the Ledoit-Wolf estimator \citep{LeWo04-estimlargecovmat03}, an estimator that uses factor models \citep{ZhPa24-covFactorModel}, where the number of factors was chosen to be equal to the number of variables, and the glasso estimator, an estimator for a sparse precision matrix, implemented in \cite{Ga18-CVglassoPackage}, where the hyperparameter is chosen using cross validation.

\paragraph*{\textbf{Performance evaluation}}
To compare the performances of the different estimators, we compare their mean absolute error (MAE) evaluated on the scale of the correlation matrix: 
\begin{equation}
\frac{1}{d^2}    \sum_{i=1}^d \sum_{j=1}^d |R_{i,j}^\star-\Bar{R}_{i,j}|,
\end{equation}
where $R^\star$ denotes the true correlation matrix and $\Bar{R}$ denotes the estimated correlation matrix.

\;

\;

\subsection{Settings within the model assumptions \label{subsec: Performance in settings with different covariance structures}}
\paragraph*{\textbf{Description}}
In this section, we compare the performance of the different estimators within the model assumption, in both \textbf{FSS} with $d=200$ and \textbf{TFR} with $d=195$.
We also study the impact of external information on the parameters. 
 
\paragraph*{\textbf{Results}}

\begin{figure}
\centering
\begin{tabular}{ccc}
\textbf{FSS} & \textbf{Known}& \textbf{Unknown}\\
\textbf{TFR} &  \includegraphics[width=162.4pt]{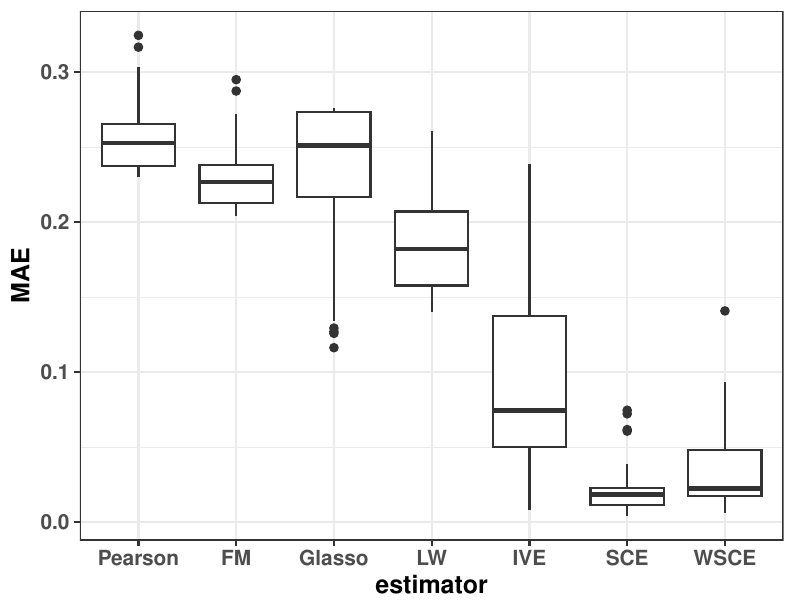}   & \includegraphics[width=162.4pt]{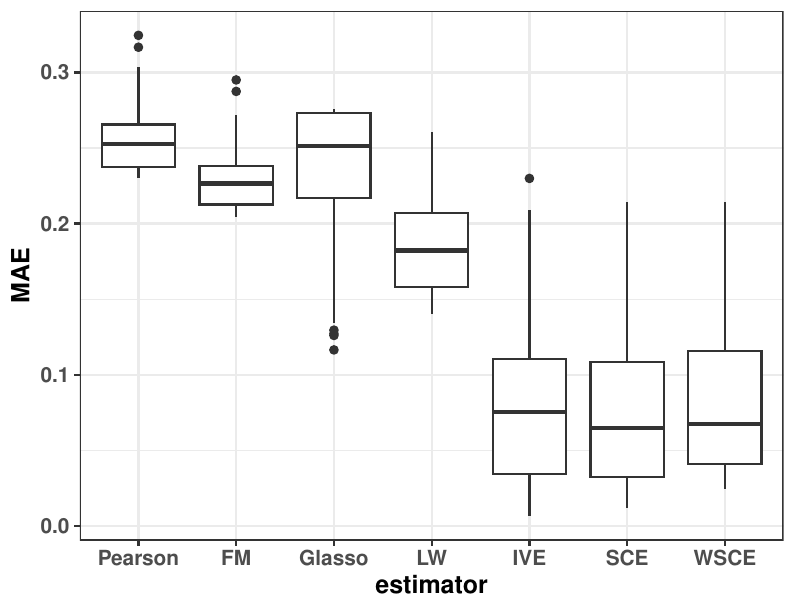}\\&\textbf{Known} &\textbf{Unknown} \\
 &  \includegraphics[width=162.4pt]{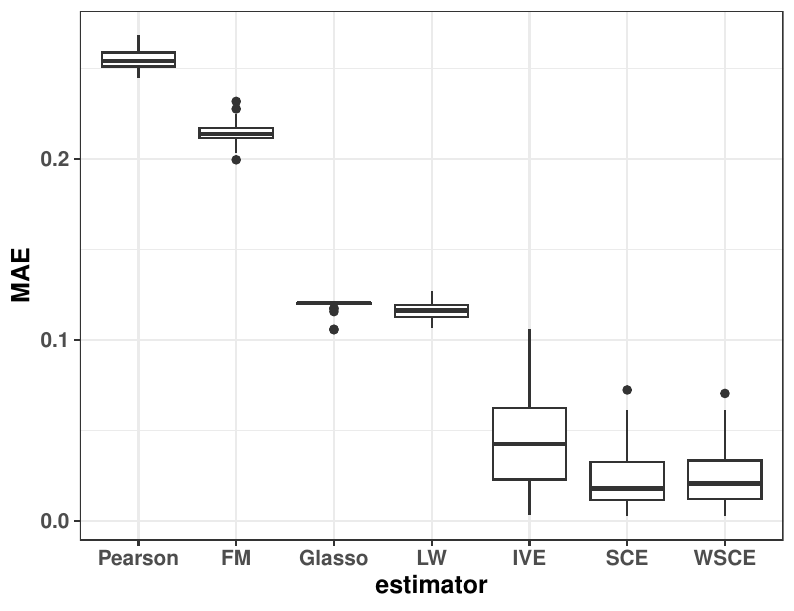}   & \includegraphics[width=162.4pt]{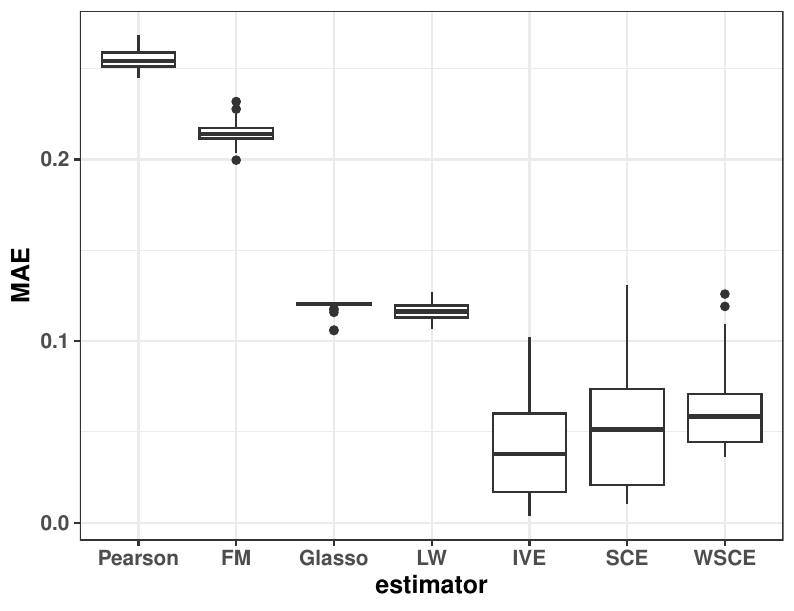}
\end{tabular}
\caption{{  Boxplots of the mean absolute error (MAE) for the Pearson correlation matrix (Pearson), the estimator that uses factor models (FM), the glasso estimator (Glasso), the Ledoit-Wolf estimator (LW), and the IVE, SCE, and WSCE for 40 independent simulations. Estimators were evaluated in the fully simulated setting (FSS) and the setting of the TFR dataset (TFR). \textbf{Left:} the case when the means and variances are known. \textbf{Right:} the case when they are unknown and estimated by using the sample mean and variance. The errors of the IVE, SCE and WSCE are the lowest, with the SCE and WSCE outperforming the IVE when the means and variances are known.
} }\label{fig: sim_0102_errors}
\end{figure}

 The results are shown in Figure \ref{fig: sim_0102_errors}. 
 In all settings, the WSCE, SCE and the IVE outperformed the other estimators. 
 Since these three estimators are the only ones that can take advantage of knowing $\mu$ and $\sigma$, they are the only ones that change between the known and unknown cases. Knowing the parameters has little impact on the IVE but improves the performances of both the SCE and WSCE. Thus, they outperformed the IVE in such a scenario. 
 Note that, whenever the mean and variance parameters are unknown, the simple version of $\hat{\lambda}_{\text{WSCE}}$, $\hat{\lambda}_{\text{WSCE}}^U$, was unstable. Thus we used the parametric bootstrap to calculate $\hat{\lambda}_{\text{WSCE}}$ (see Appendix C for more details). 
 
\subsection{Performance with varying dimensions of the data\label{subsec: Performance with varying dimensions}}
We now describe the performance of the WSCE for different values of $d$. 

\paragraph*{\textbf{Description}} We consider the \textbf{TFR} setting with known mean and variance parameters. We selected subsets of the data of sizes $d=14,32,65,115,195$. These subsets are created by cumulatively including countries in Southern and Middle Africa, Eastern Africa, Western and Northern Africa, Asia and America, and all other countries, respectively.

\begin{figure}
\centering
\includegraphics[width=406.0pt]{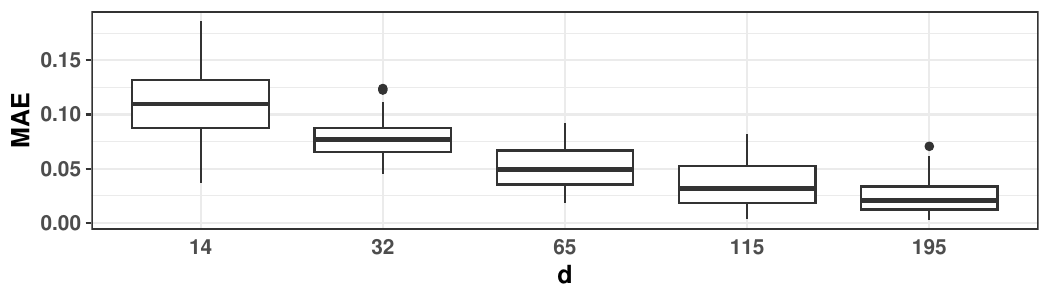}
\caption{{Boxplot of the mean absolute error (MAE) of the WSCE with different values of $d$, repeated for 40 independent simulations. The values of $d$ correspond to the number of countries in the following regions (added cumulatively): Southern and Middle Africa, Eastern Africa, Western and Northern Africa, Asia and America, the remaining countries in the dataset.}}\label{fig: sim_02_different_n}
\end{figure} 

\paragraph*{\textbf{Results}} The results are shown in Figure \ref{fig: sim_02_different_n}.  $T=11$ was fixed, and $40$ different datasets were simulated independently for each value of $d$.  The mean absolute error decreases with the number of countries, $d$. This was expected due to the consistency in $d$ of the parameters (Theorem \ref{thm-consistent_in_n}). In other words, adding more countries to the dataset improved the performance of the WSCE.

\subsection{Performance under model misspecification and the presence of missing values\label{subsec: Performance under model misspecification}}

\paragraph*{\textbf{Settings}}  
Here, we study the impact of model misspecification on our estimators. We replace $R$ by $R_{\text{miss}}$ in Equation~\eqref{eq :simu}, where 
\begin{align*}
    R_{miss}&=\xi R(\alpha,\beta,\delta)+(1-\xi)\tilde{R},
\end{align*}
with $\xi \in [0,1]$,  $R(\alpha,\beta,\delta)$ denoting the correlation structure from the \textbf{TFR} setting and $\tilde{R}$ being simulated using an additional matrix $F_{miss}$ from the \textbf{FSS} setting, which is not given to the model.
Thus, the model assumptions hold when $\xi=0$, and the correlation structure does not depend on the observed covariates  when $\xi=1$.

In the TFR dataset, the values of the standardized errors are missing for countries that have not yet entered phase II or III of the model of \cite{Al_et_al11-tfr_projections}. Thus, in this scenario, we consider two settings: one without missing values and the other one with missing values. In the latter scenario, we set the values of $Y_t$ that were missing in the TFR dataset to be  missing. The IVE, SCE and WSCE need to be adapted under the presence of missing values. The corresponding procedure is described in Appendix E.
    
\begin{figure}
    \centering
     \includegraphics[width=406.0pt]{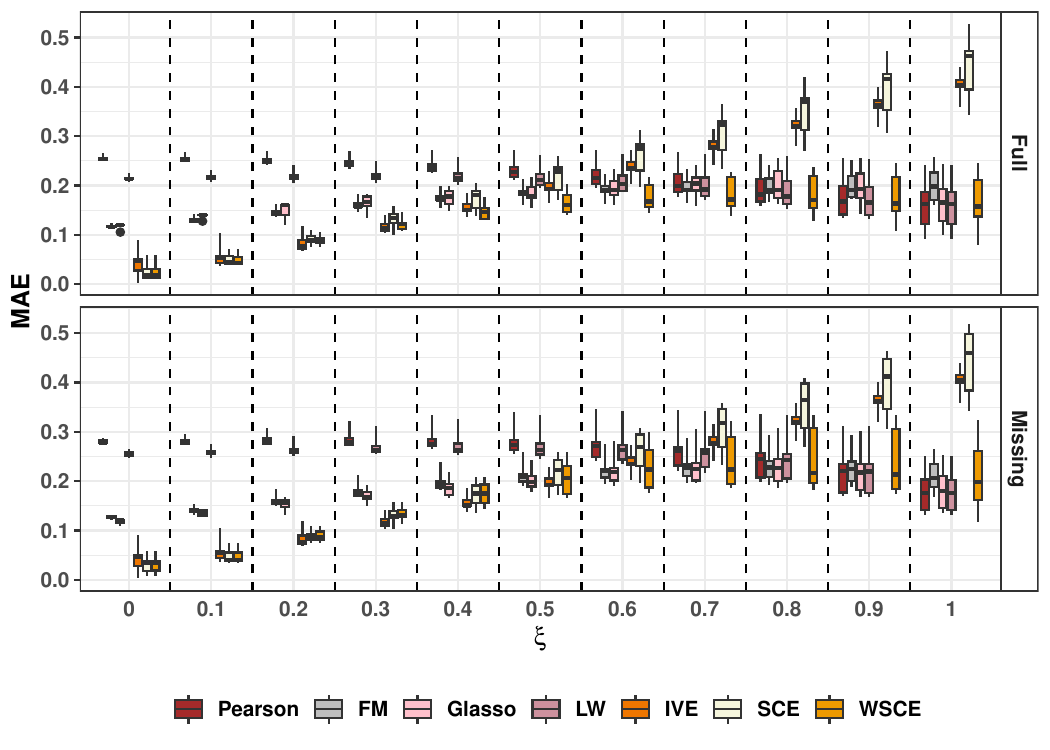} 
    \caption{Boxplots of the mean absolute error (MAE) of the Pearson correlation matrix (Pearson), the Ledoit-Wolf estimator (LW), the estimator that uses factor models (FM), the glasso estimator (Glasso), IVE, SCE and WSCE, for $\xi=0,0.1,\dots,0.9,1$. 10 independent simulations were calculated by adding additional structure to the correlation, which is unknown and randomly simulated. At $\xi=1$, the correlation structure does not depend on the observed covariates. Full: no missing values. Missing: values that were missing in the original dataset were also set to be missing. The results do not change much between these two settings. This indicates that the missing-value imputation used is quite robust. \label{fig:modelmisspec_combined}}
\end{figure}

        \paragraph*{\textbf{Results}} The results are shown in Figure \ref{fig:modelmisspec_combined}. Ten independent datasets were simulated for each value of $\xi$ and estimators were evaluated on the scale of the correlation matrix. The SCE and the WSCE perform similarly and outperform the other estimators in the scenarios that are not too far away from the model assumptions. 
        As expected, the performances of both the SCE and the WSCE deteriorate when the value of $\xi$ increases, meaning that the scenario gets further away from the model assumptions. However, when the scenario is very different from the model assumptions ($\xi > 0.5$), the WSCE outperforms the SCE and performs at least as well as the other estimators.

\section{Covariance estimates for the TFR dataset\label{sec: Covariance estimates of the TFR dataset}}
In this section, we will study the TFR dataset. 
We start by describing the data in detail. Then, we calculate our estimators with and without interaction terms, and  we finish by performing model selection. 
%\subsection{The whole world}
% Fosdick and Raftery (without missing values and with missing values)
\subsection{TFR data description}

As described in Section~\ref{subsec: The total fertility rate covariance study}, we want to estimate the covariance matrix of the total fertility rate (TFR) for 195 countries. Since we are in a Markov model with dependent observations, we estimate the covariance matrix of the TFR conditional on its preceding values, where the covariance matrix can vary with time. We assume that we are in the model described by Equation ~\eqref{epsilon_model}. 

Let us recall that in this model, the data $(Y_1,\dots, Y_T)$ denotes the TFR at time $1,\dots, T$. We fit the model described in Equation~\eqref{eq:log_lik} with 
\begin{equation}
 \label{eq: mod_withoutinter}  
R(\alpha,\beta,\delta) = \alpha_A F_A +\alpha_B F_B + \alpha_C F_C + \delta_D\Gamma(\beta_D)^{-1}+\alpha_EI_d,
\end{equation}
where 
$(F_A)_{i,j},(F_B)_{i,j}$ are equal to 1 if country $i$ and $j$ have the same common colonizer or belong to the same UN region, respectively, $(F_C)_{i,j}$ is always 1, and $\Gamma(\beta_D)$ is described by Equation~\eqref{eq:gamma}.

{\begin{table}[t!]
    \centering
    \caption{{The number of countries in phase I of the model of \cite{Al_et_al11-tfr_projections} for each of the successive 5-year periods used in the TFR dataset. They are decreasing from the period 1950-1955, during which all countries were in phase I, to the period 2005-2010, during which all countries are no longer in phase I.}}

    \;
    
    \begin{tabular}{c|c|c|c|c|c}
1950-1955&1955-1960&1960-1965&1965-1970&1970-1975&1975-1980\\196&121&98&74&56&35\\\hline1980-1985&1985-1990&1990-1995&1995-2000&2000-2005&2005-2010\\26&7&4&2&0&0
    \end{tabular}
    
    \label{tab:phaseI_countries}
\end{table}}

\paragraph*{\textbf{Selected countries}}
Vanuatu was reportedly colonized by two countries. This non-unique cluster membership could be modelled by splitting the common colonizer covariate into multiple distinct covariates for each respective cluster. However, this would unreasonably increase the number of parameters that we would need to estimate. Thus, we removed the corresponding variable and worked with the remaining $d=195$ countries.

\paragraph*{\textbf{Missing values}}
We aim at estimating the covariance matrix for country pairs where both countries have entered either phase II or III of the model of \cite{Al_et_al11-tfr_projections}.
The TFR values of the countries that are still in phase I are thus treated as missing values. 
As time passes, more countries go from phase I to phase II. Thus,  the number of missing values changes between the observations.
We give the  number of countries that are in phase I at each time period in Table \ref{tab:phaseI_countries}.  
The  values of the standardized errors $\varepsilon_{t,j}$, that are used to estimate the covariance matrix, are assumed to be missing at random everywhere. Thus, covariance estimation is appropriate on the marginal distribution of the non-missing values \citep{Ru76-missing_at_random,Sh_et_al13-missingdata}.

\begin{remark}
Once a country leaves phase I, it will no longer return to it. Thus, the missing value structure is monotone and by construction the values of $\varepsilon_{t,j}$ are missing at random everywhere because the observed data vector directly implies the positions of the missing and observed data. 
\end{remark}

\subsection{ Covariance without interaction}
In this section we compute the SCE by estimating the parameters of Equation~\eqref{eq: mod_withoutinter}. 

\paragraph*{\textbf{Impact of the covariate}}
We introduce the concept of average effects to compare the effects of the different covariates in the model. The average effect of a covariate is the average correlation in the data that is due to this covariate. For direct effects, this corresponds to the value of the linear coefficients ($"$common colonizer$"$, $"$same region$"$). For the contiguity effect, we take the overall mean effect of country pairs which are direct neighbors of each other,
$$
\eta_{contig} =\frac{1}{\sum_{i\neq j}M_{i,j}}  \sum_{i\neq j\text{{ s.t. }} M_{i,j}=1} \delta_D \Gamma(\beta_D)_{i,j}^{-1}.
$$
\begin{remark}
$\eta_{contig}$ is not equal to $\delta_D$ because, contrary to the values of $F_A$ or $F_B$, $\Gamma(\beta_D)_{i,j}^{-1}$ is not always equal to $0$ or $1$.
\end{remark}
The rationale is that if one adds all the pertinent coefficients for a given covariate, one gets the mean of the correlations for data points that have this covariate in common. For instance: 
$$
\alpha_B + \eta_{contig}, 
$$
gives the mean of the estimated correlations of countries that are neighbors and in the same region, but not with the same colonizer.

The estimated average effects are given in Table \ref{tab:result_base_model}. 

\subsection{Interaction effects and model selection}
We can see in Table~\ref{tab:result_base_model} that at least two effects needed to overlap for countries to have a correlation higher than 0.2. This was not the case in \cite{FoRa14-corlinearreg}, where, e.g., the contiguity effect alone accounted for a correlation of 0.26 for all country pairs with TFRs below 5. 
We wanted to check if there were interaction effects. Indeed the neighborhood effect, for instance, may be different if you are in the same region or not. 

\begin{figure}
\centering
\includegraphics[width=406.0pt]{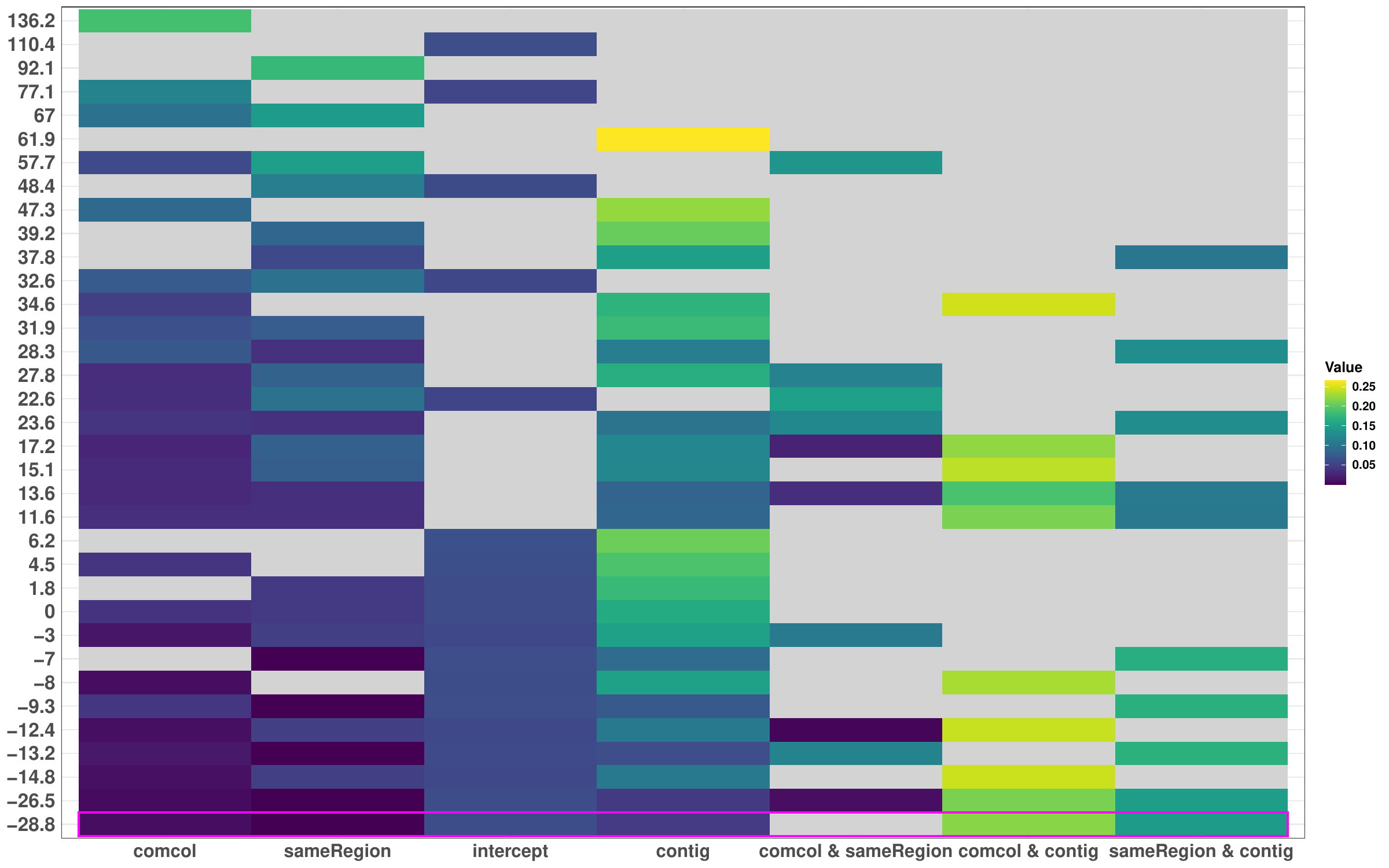}
\caption{{On the $y$-axis: values of the BIC for all 35 models tested; the BIC is centered such that the model which includes all but interaction effects shows a value of 0; grey squares indicate effects that are not included in their respective model; on the $x$-axis: different average effects; we can see that only models which include interaction effects have a lower BIC than this model; the model with the lowest BIC set only the combined effect of the $"$regional$"$ and $"$common-colonizer$"$ covariates to 0.}}\label{fig: modelchoice}
\end{figure}

\begin{figure}
\centering
\includegraphics[width=406.0pt]{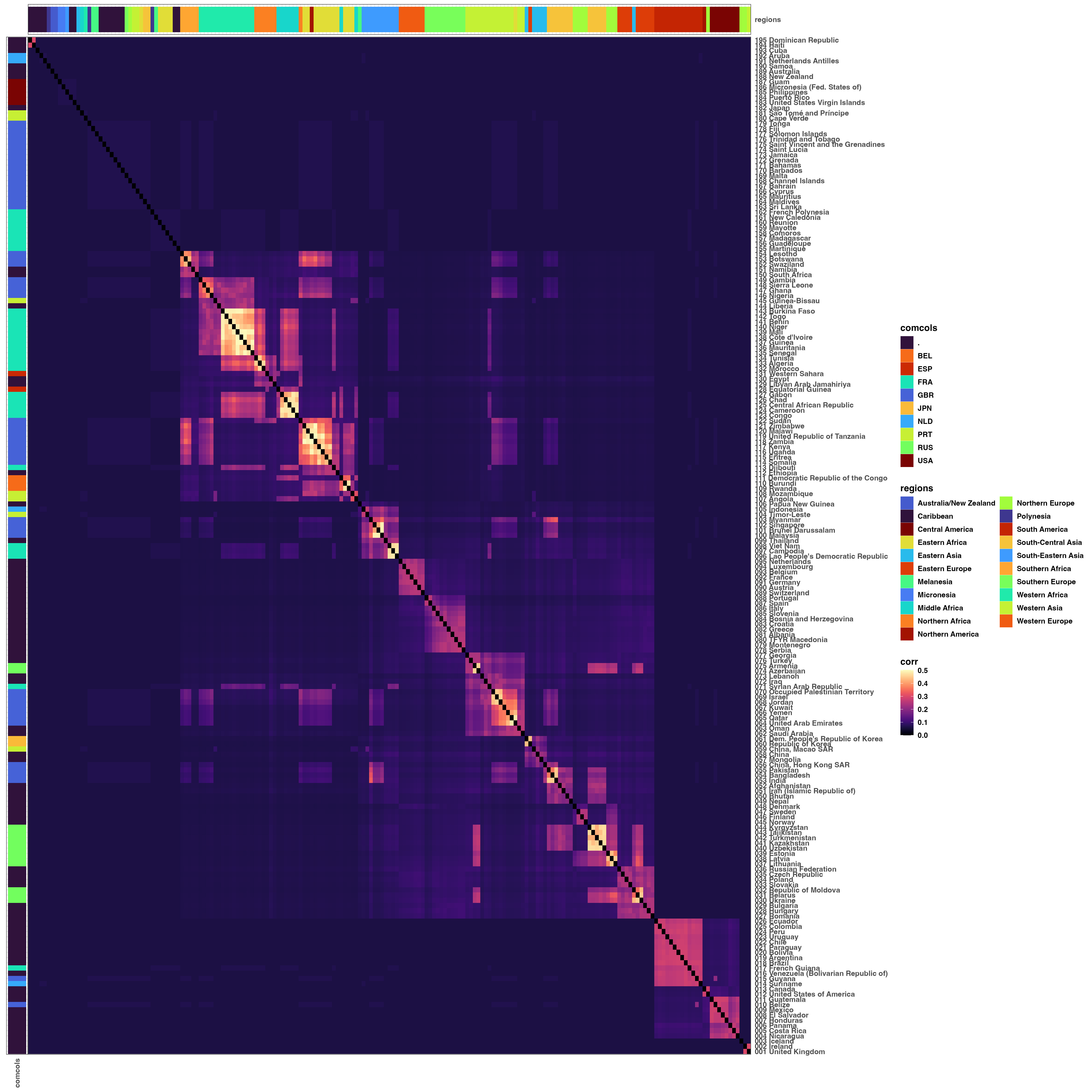}
\caption{{A heatmap of the correlation matrix calculated from the WSCE (diagonal entries were set to 0 to improve the visualization); on the x-axis: colonizers (countries with no colonizer were set to belong to colonizer $"$.$"$); on the y-axis: regions; Countries are particularly correlated if they are close to each other and are either in the same region or share the same common colonizer. The matrix was sorted using the order given by hierarchical clustering with the Ward distance on the dissimilarity matrix $1-\hat{R}_{\text{WSCE}}$, using the option "$h$-clust" in the "corrplot" package \citep{corrplot2021}.
}}\label{fig: heatmaps}
\end{figure}

{\begin{table}[t!]
    \centering
    \caption{{ Average effects of the model in which all effects are included. All effects were rounded to the third decimal place.}}

    \;
    
    \begin{tabular}{c|c|c|c|c|c}
    &comcol&sameRegion&intercept&contig&\\
    \hline &0.038&0.044&0.06&0.162&\\\end{tabular}
    
    \label{tab:result_base_model}
\end{table}}

Thus, in addition to the intercept, $"$common colonizer$"$, $"$same region$"$ and the spatial effect we add their interactions by adding random effects with correlation matrices equal to the Hadamard product of the correlation matrices of the individual effects. \citet{Ma_et_al20-covmatHadamardProduct} do that with covariance matrices but since we separate the variance from the correlation matrix estimation, we use correlation matrices instead. 

We can include up to 3 interaction effects, with correlation matrices\begin{align*}
    &F_{A,B}=F_A\odot F_B,\text{ common colonizer and same region effect},\\&F_{A,D}=F_A\odot\Gamma(\beta_D)^{-1},\text{ common colonizer and spatial effect,}\\&F_{B,D}=F_B\odot\Gamma(\beta_D)^{-1},\text{ same region and spatial effect}.
\end{align*}

This gives up to 8 effects. We select which effect we should include by computing the BIC for each of the possible $128=2^8$ models. However, we exclude interaction effects whenever one of their individual component effects is excluded. This reduces the scope of our model selection to 35 models, the results of which are plotted in Figure \ref{fig: modelchoice}. 

The BIC was centered in the base model described in Equation~\eqref{eq: mod_withoutinter}. Interestingly, the only models with a better (lower) BIC than the base model are the ones that do include interaction effects. This is in agreement with~\cite{FoRa14-corlinearreg}, which kept all these effects, but did not try to add interactions.  
The model with the lowest BIC is the model that includes all but one interaction effect, the effect that accounts for the interaction between $"$common colonizer$"$ and the $"$same region$"$ effect. 

Just as for the previous model, we compared the average effects in this model. For interactions between direct effects (common colonizer and same region), the average effect is the coefficient $\alpha_{A,B}$. 
 For interaction effects that involve the contiguity effect, we take the mean effect of all pairs of countries that are neighbors and in the same region (resp., have the same colonizer):
$$
\eta_{contig,B} =\frac{1}{\sum_{i\neq j}(F_B\odot M)_{i,j}}  \sum_{i\neq j\text{{ s.t. }} M_{i,j}=1 \;\&\;  (F_{B})_{i,j}=1 } \alpha_{B,D} (F_B\odot\Gamma(\hat{\beta}_{\text{SCE}})^{-1})_{i,j}.
$$
{\begin{table}[t!]
    \centering
    \caption{{ Average effects of the model chosen in Figure \ref{fig: modelchoice}, in which all effects and their interactions, but the interaction of the $"$common colonizer$"$ and the $"$same region$"$ effect are included. All effects but the regional effect were rounded to the third decimal place.}}
    \begin{tabular}{c|c|c|c|c|c|c|c}
    &comcol&sameRegion&intercept&contig&comcol and contig&sameRegion and contig&\\ \hline &0.008&2e-06&0.061&0.045&0.218&0.146 &\\ 
    \end{tabular}
    
    \label{tab:result_interaction_effects}
\end{table}}

Table~\ref{tab:result_interaction_effects} gives the average effects of the selected model. Effects are much higher when two attributes overlap.

The correlation matrix obtained with these coefficients is plotted in Figure \ref{fig: heatmaps}.
Except for some clusters of countries, the TFRs of two countries are mostly estimated to not be highly correlated, given their previous TFRs. 

To check if we do not miss correlations that may come from an effect that is not included in the model, we computed the WSCE. However, it was equal to the SCE. In our simulation settings this corresponds to the case where our model assumptions were correct. Thus, we can find no evidence against our model assumptions.

%\subsection{Results}

\section{Discussion}

We introduced the structured covariance estimator (SCE) and the weighted structured covariance estimator (WSCE),  estimators for large covariance matrices in the presence of pairwise covariates. We showed consistency and asymptotic normality of these estimators in the dimension of the data and in the number of data points and gave a procedure for estimating their confidence regions, under mild assumptions. Furthermore, we tested our estimators in scenarios in which some part of our model was misspecified, where the WSCE performed well.

We incorporated pairwise information into the covariance matrix estimation by modelling the standardized errors as a sum of weighted and standardized random effects. A very different approach from ours would be to penalize the estimator of the covariance matrix directly. This was done by \citet{Li_et_al14-thresholding_covmat}, and in a Bayesian way by \cite{AzRa18-estimalargecovmat01}, both of which use a weight matrix to penalize individual matrix entries. However, this requires direct estimation of all parameters of the covariance matrix, of which there are $195 \times 194/2 = 18,915$ in our case. We do not need to do this in our model, where we only require efficient estimation of a small number of parameters. In fact, \cite{AzRa18-estimalargecovmat01} pointed out that the high dimension of their parameter space made it impractical for them to carry out an MCMC simulation in their Bayesian setting. In contrast, an extension of our method to the Bayesian paradigm would be straightforward, since we only need to simulate a vector of dimension 6.

The initial value of our estimator, the IVE, performed well in our simulation study. The WSCE is an asymptotically optimal interpolation between the SCE and the Pearson correlation matrix, as long as the underlying distribution of the data is Gaussian. However, if the distribution is very different from a Gaussian distribution, it might perform suboptimally. In this case, one could instead use a linear interpolation between the IVE and the Pearson correlation matrix, since the IVE is consistent and asymptotically normal if our correlation structure holds, but the data are non-Gaussian.

There is also the question of whether we should be using the Pearson correlation matrix in constructing the WSCE. We obtained decent results even in scenarios far from our model assumptions, but  note that the Pearson correlation matrix is known to not behave well in  settings with small sample size and a high number of variables. 
Depending on the setting, one can decide to use a more adapted estimator in the construction of the WSCE, such as the Ledoit-Wolf estimator or Glasso for instance.  

A similar argument can be made for our mean- and variance estimators: we used the fact that accurate estimates of the mean and variance of the data were already provided. This is convenient. but not necessary. In cases where these estimates are not provided we recommend estimating the mean-, variance-, and correlation structure jointly if computationally feasible.

Finally, the WSCE could be adapted to very general settings such as generalized linear models (GLMs), in a similar vein to \cite{BoJo16-MCGLM}. 

The name of the WSCE refers to the fact that we have a weighted combination of correlation structures defined via known covariates. It is not to be confused with techniques that have similar names, but only focus on the estimation of one given matrix structure, such as \citet{Bu_et_al82-same_name_02,Su_et_al16-same_name_04,Lo_et_al23-same_name_03}.

\medskip

\textbf{Acknowledgements} The authors would like to thank Daniel Suen for recommending the re-use of the initialization step of the estimator in the bootstrapping procedure (instead of reevaluating it in each sample that was simulated), and Julie Josse for her recommendation of \cite{JoHu16-missingvalueimputation}. 

\begin{funding}
Raftery's research was supported by NIH grant R01 HD-070936, the Fondation des Sciences Math\'{e}matiques de Paris (FSMP), and Universit\'{e} Paris Cit\'{e} (UPC). He thanks the Laboratoire MAP5 at UPC for warm hospitality.
\end{funding}

\begin{supplement}
\stitle{Supplementary material}
\sdescription{Contains Appendix A (derivatives of the likelihood), Appendix B (proofs), Appendix C (details on the algorithm), Appendix D (alternative model justifications), and Appendix E (simulation settings details).}
\end{supplement}

\bibliographystyle{chicago}
\bibliography{biblio.bib}

\begin{thebibliography}{}

\bibitem[\protect\citeauthoryear{Aguilar and West}{Aguilar and West}{2000}]{AgWe00-BayesianFactorModels02}
Aguilar, O. and M.~West (2000).
\newblock Bayesian dynamic factor models and portfolio allocation.
\newblock {\em Journal of Business \& Economic Statistics\/}~{\em 18\/}(3), 338--357.

\bibitem[\protect\citeauthoryear{Alkema, Raftery, Gerland, Clark, Pelletier, Buettner, and Heilig}{Alkema et~al.}{2011a}]{Al_et_al11-tfr_projections}
Alkema, L., A.~E. Raftery, P.~Gerland, S.~J. Clark, F.~Pelletier, T.~Buettner, and G.~K. Heilig (2011a).
\newblock Probabilistic projections of the total fertility rate for all countries.
\newblock {\em Demography\/}~{\em 48\/}(3), 815--839.

\bibitem[\protect\citeauthoryear{Alkema, Raftery, Gerland, Clark, Pelletier, Buettner, and Heilig}{Alkema et~al.}{2011b}]{Al_et_al11-online_resource_01}
Alkema, L., A.~E. Raftery, P.~Gerland, S.~J. Clark, F.~Pelletier, T.~Buettner, and G.~K. Heilig (2011b).
\newblock Probabilistic projections of the total fertility rate for all countries, {O}nline {R}esource 1.
\newblock {\em Demography\/}~{\em 48\/}(3), 815--839.

\bibitem[\protect\citeauthoryear{Anderson}{Anderson}{1973}]{An73-firstlinearcovarpaper}
Anderson, T.~W. (1973).
\newblock Asymptotically efficient estimation of covariance matrices with linear structure.
\newblock {\em The Annals of Statistics\/}~{\em 1\/}(1), 135--141.

\bibitem[\protect\citeauthoryear{Azose and Raftery}{Azose and Raftery}{2018}]{AzRa18-estimalargecovmat01}
Azose, J.~J. and A.~E. Raftery (2018).
\newblock Estimating large correlation matrices for international migration.
\newblock {\em The annals of applied statistics\/}~{\em 12\/}(2), 940.

\bibitem[\protect\citeauthoryear{Barnard, McCulloch, and Meng}{Barnard et~al.}{2000}]{Ba_et_al00-separationstrategy01}
Barnard, J., R.~McCulloch, and X.-L. Meng (2000).
\newblock Modeling covariance matrices in terms of standard deviations and correlations, with application to shrinkage.
\newblock {\em Statistica Sinica\/}~{\em 10\/}(4), 1281--1311.

\bibitem[\protect\citeauthoryear{Bates, Maechler, and Jagan}{Bates et~al.}{2024}]{nearPD_package}
Bates, D., M.~Maechler, and M.~Jagan (2024).
\newblock {\em Matrix: Sparse and Dense Matrix Classes and Methods}.
\newblock R package version 1.7-0.

\bibitem[\protect\citeauthoryear{Bernardo, Bayarri, Berger, Dawid, Heckerman, Smith, and West}{Bernardo et~al.}{2003}]{Be_et_al03-BayesianFactorModels}
Bernardo, J., M.~Bayarri, J.~Berger, A.~Dawid, D.~Heckerman, A.~Smith, and M.~West (2003).
\newblock Bayesian factor regression models in the “large p, small n” paradigm.
\newblock {\em Bayesian statistics\/}~{\em 7}, 733--742.

\bibitem[\protect\citeauthoryear{Besag and Kooperberg}{Besag and Kooperberg}{1995}]{BeKo95-CAR_GMRF_01}
Besag, J. and C.~Kooperberg (1995).
\newblock On conditional and intrinsic autoregressions.
\newblock {\em Biometrika\/}~{\em 82\/}(4), 733--746.

\bibitem[\protect\citeauthoryear{Besag, York, and Molli{\'e}}{Besag et~al.}{1991}]{Be_et_al91-CAR_GMRF_02}
Besag, J., J.~York, and A.~Molli{\'e} (1991).
\newblock Bayesian image restoration, with two applications in spatial statistics.
\newblock {\em Annals of the institute of statistical mathematics\/}~{\em 43}, 1--20.

\bibitem[\protect\citeauthoryear{Bonat and J{\o}rgensen}{Bonat and J{\o}rgensen}{2016}]{BoJo16-MCGLM}
Bonat, W.~H. and B.~J{\o}rgensen (2016).
\newblock Multivariate covariance generalized linear models.
\newblock {\em Journal of the Royal Statistical Society Series C: Applied Statistics\/}~{\em 65\/}(5), 649--675.

\bibitem[\protect\citeauthoryear{Broyden}{Broyden}{1970}]{Br70-BFGS02}
Broyden, C.~G. (1970).
\newblock The convergence of a class of double-rank minimization algorithms 1. general considerations.
\newblock {\em IMA Journal of Applied Mathematics\/}~{\em 6\/}(1), 76--90.

\bibitem[\protect\citeauthoryear{Burg, Luenberger, and Wenger}{Burg et~al.}{1982}]{Bu_et_al82-same_name_02}
Burg, J.~P., D.~G. Luenberger, and D.~L. Wenger (1982).
\newblock Estimation of structured covariance matrices.
\newblock {\em Proceedings of the IEEE\/}~{\em 70\/}(9), 963--974.

\bibitem[\protect\citeauthoryear{Cavazani~de Freitas, de~Oliveira~Carlos, Ligocki~Campos, and Bonat}{Cavazani~de Freitas et~al.}{2022}]{Ca_et_al22-transitiveattributes}
Cavazani~de Freitas, L.~A., L.~de~Oliveira~Carlos, A.~C. Ligocki~Campos, and W.~H. Bonat (2022).
\newblock Hypothesis tests for multiple responses regression: effect of probiotics on addiction and binge eating disorder.
\newblock {\em arXiv e-prints\/}, arXiv--2208.

\bibitem[\protect\citeauthoryear{Cheng and Higham}{Cheng and Higham}{1998}]{ChHi98-nearPDalgorithm}
Cheng, S.~H. and N.~J. Higham (1998).
\newblock A modified {C}holesky algorithm based on a symmetric indefinite factorization.
\newblock {\em SIAM Journal on Matrix Analysis and Applications\/}~{\em 19\/}(4), 1097--1110.

\bibitem[\protect\citeauthoryear{Chiu, Leonard, and Tsui}{Chiu et~al.}{1996}]{Ch_et_al96-covmatLoglink}
Chiu, T.~Y., T.~Leonard, and K.-W. Tsui (1996).
\newblock The matrix-logarithmic covariance model.
\newblock {\em Journal of the American Statistical Association\/}~{\em 91\/}(433), 198--210.

\bibitem[\protect\citeauthoryear{Christensen and Amemiya}{Christensen and Amemiya}{2003}]{ChAm03-multivariate_spatial_factoranalysis}
Christensen, W.~F. and Y.~Amemiya (2003).
\newblock Modeling and prediction for multivariate spatial factor analysis.
\newblock {\em Journal of Statistical planning and inference\/}~{\em 115\/}(2), 543--564.

\bibitem[\protect\citeauthoryear{Erdős and Rényi}{Erdős and Rényi}{1959}]{ErRe1959randomgraphs}
Erdős, P. and A.~Rényi (1959).
\newblock On random graphs i.
\newblock {\em Publ. Math.\/}~{\em 6\/}(290-297), 18.

\bibitem[\protect\citeauthoryear{Fan, Liao, and Liu}{Fan et~al.}{2016}]{Fa_et_al15-estimlargecovmat04}
Fan, J., Y.~Liao, and H.~Liu (2016, 03).
\newblock An overview of the estimation of large covariance and precision matrices.
\newblock {\em The Econometrics Journal\/}~{\em 19\/}(1), C1--C32.

\bibitem[\protect\citeauthoryear{Fletcher and Reeves}{Fletcher and Reeves}{1964}]{FlRe64-optimDefinitionBFGS}
Fletcher, R. and C.~M. Reeves (1964).
\newblock Function minimization by conjugate gradients.
\newblock {\em The computer journal\/}~{\em 7\/}(2), 149--154.

\bibitem[\protect\citeauthoryear{Fosdick and Raftery}{Fosdick and Raftery}{2014}]{FoRa14-corlinearreg}
Fosdick, B.~K. and A.~E. Raftery (2014).
\newblock Regional probabilistic fertility forecasting by modeling between-country correlations.
\newblock {\em Demographic Research\/}~{\em 30\/}(35), 1011.

\bibitem[\protect\citeauthoryear{Freni-Sterrantino, Ventrucci, and Rue}{Freni-Sterrantino et~al.}{2018}]{Fr_et_al18-disconnected_plus_singletons}
Freni-Sterrantino, A., M.~Ventrucci, and H.~Rue (2018).
\newblock A note on intrinsic conditional autoregressive models for disconnected graphs.
\newblock {\em Spatial and spatio-temporal epidemiology\/}~{\em 26}, 25--34.

\bibitem[\protect\citeauthoryear{Friedman, Hastie, and Tibshirani}{Friedman et~al.}{2008}]{Fr_et_al08-graphical_lasso}
Friedman, J., T.~Hastie, and R.~Tibshirani (2008).
\newblock Sparse inverse covariance estimation with the graphical lasso.
\newblock {\em Biostatistics\/}~{\em 9\/}(3), 432--441.

\bibitem[\protect\citeauthoryear{Ga{\l}ecki and Burzykowski}{Ga{\l}ecki and Burzykowski}{2013}]{GaBu13-LMM_book}
Ga{\l}ecki, A. and T.~Burzykowski (2013).
\newblock {\em Linear Mixed-Effects Models Using R: A Step-by-Step Approach}.
\newblock New York, NY: Springer New York.

\bibitem[\protect\citeauthoryear{Galloway}{Galloway}{2018}]{Ga18-CVglassoPackage}
Galloway, M. (2018).
\newblock {\em CVglasso: Lasso Penalized Precision Matrix Estimation}.
\newblock R package version 1.0.

\bibitem[\protect\citeauthoryear{Gamerman, Lopes, and Salazar}{Gamerman et~al.}{2008}]{Ga_et_al08-spatial_dynamic_factoranalysis}
Gamerman, D., H.~F. Lopes, and E.~Salazar (2008).
\newblock {Spatial dynamic factor analysis}.
\newblock {\em Bayesian Analysis\/}~{\em 3\/}(4), 759 -- 792.

\bibitem[\protect\citeauthoryear{Goldfarb}{Goldfarb}{1970}]{Go70-BFGS03}
Goldfarb, D. (1970).
\newblock A family of variable-metric methods derived by variational means.
\newblock {\em Mathematics of Computation\/}~{\em 24\/}(109), 23--26.

\bibitem[\protect\citeauthoryear{Goldfarb and Idnani}{Goldfarb and Idnani}{1983}]{GoId83-quadratic_optimization02}
Goldfarb, D. and A.~Idnani (1983).
\newblock A numerically stable dual method for solving strictly convex quadratic programs.
\newblock {\em Mathematical programming\/}~{\em 27\/}(1), 1--33.

\bibitem[\protect\citeauthoryear{Goldfarb and Idnani}{Goldfarb and Idnani}{2006}]{GoId06-quadratic_optimization01}
Goldfarb, D. and A.~Idnani (2006).
\newblock Dual and primal-dual methods for solving strictly convex quadratic programs.
\newblock In {\em Numerical Analysis: Proceedings of the Third IIMAS Workshop Held at Cocoyoc, Mexico, January 1981}, pp.\  226--239. Springer.

\bibitem[\protect\citeauthoryear{Hannan}{Hannan}{1973}]{Ha73-asconvergence_timeseries_params}
Hannan, E.~J. (1973).
\newblock The asymptotic theory of linear time-series models.
\newblock {\em Journal of Applied Probability\/}~{\em 10\/}(1), 130--145.

\bibitem[\protect\citeauthoryear{Harshman and Lundy}{Harshman and Lundy}{1994}]{HaLu94-parallel_factor_analysis}
Harshman, R.~A. and M.~E. Lundy (1994).
\newblock Parafac: Parallel factor analysis.
\newblock {\em Computational Statistics \& Data Analysis\/}~{\em 18\/}(1), 39--72.

\bibitem[\protect\citeauthoryear{Josse and Husson}{Josse and Husson}{2016}]{JoHu16-missingvalueimputation}
Josse, J. and F.~Husson (2016).
\newblock {missMDA}: A package for handling missing values in multivariate data analysis.
\newblock {\em Journal of Statistical Software\/}~{\em 70\/}(1), 1--31.

\bibitem[\protect\citeauthoryear{Karolyi}{Karolyi}{1992}]{Ka92-BayesianCorrEst_withGroups02}
Karolyi, G.~A. (1992).
\newblock Predicting risk: Some new generalizations.
\newblock {\em Management Science\/}~{\em 38\/}(1), 57--74.

\bibitem[\protect\citeauthoryear{Karolyi}{Karolyi}{1993}]{Ka93-BayesianCorrEst_withGroups}
Karolyi, G.~A. (1993).
\newblock A {B}ayesian approach to modeling stock return volatility for option valuation.
\newblock {\em Journal of Financial and Quantitative Analysis\/}~{\em 28\/}(4), 579--594.

\bibitem[\protect\citeauthoryear{Kyung and Ghosh}{Kyung and Ghosh}{2010}]{Ky_et_al10-DCARmodel}
Kyung, M. and S.~K. Ghosh (2010).
\newblock Maximum likelihood estimation for directional conditionally autoregressive models.
\newblock {\em Journal of Statistical Planning and Inference\/}~{\em 140\/}(11), 3160--3179.

\bibitem[\protect\citeauthoryear{Ledoit and Wolf}{Ledoit and Wolf}{2003}]{LeWo03-CovMatCombination}
Ledoit, O. and M.~Wolf (2003).
\newblock Improved estimation of the covariance matrix of stock returns with an application to portfolio selection.
\newblock {\em Journal of empirical finance\/}~{\em 10\/}(5), 603--621.

\bibitem[\protect\citeauthoryear{Ledoit and Wolf}{Ledoit and Wolf}{2004}]{LeWo04-estimlargecovmat03}
Ledoit, O. and M.~Wolf (2004).
\newblock A well-conditioned estimator for large-dimensional covariance matrices.
\newblock {\em Journal of multivariate analysis\/}~{\em 88\/}(2), 365--411.

\bibitem[\protect\citeauthoryear{Ledoit and Wolf}{Ledoit and Wolf}{2022}]{LeWo22-estimlargecovmat02}
Ledoit, O. and M.~Wolf (2022).
\newblock The power of (non-) linear shrinking: A review and guide to covariance matrix estimation.
\newblock {\em Journal of Financial Econometrics\/}~{\em 20\/}(1), 187--218.

\bibitem[\protect\citeauthoryear{Lewandowski, Kurowicka, and Joe}{Lewandowski et~al.}{2009}]{Le_et_al09-LKJprior}
Lewandowski, D., D.~Kurowicka, and H.~Joe (2009).
\newblock Generating random correlation matrices based on vines and extended onion method.
\newblock {\em Journal of Multivariate Analysis\/}~{\em 100\/}(9), 1989--2001.

\bibitem[\protect\citeauthoryear{Liechty, Liechty, and M{\"u}ller}{Liechty et~al.}{2004}]{Li_et_al04-BayesianCorrEsts}
Liechty, J.~C., M.~W. Liechty, and P.~M{\"u}ller (2004).
\newblock Bayesian correlation estimation.
\newblock {\em Biometrika\/}~{\em 91\/}(1), 1--14.

\bibitem[\protect\citeauthoryear{Liu, Wang, and Zhao}{Liu et~al.}{2014}]{Li_et_al14-thresholding_covmat}
Liu, H., L.~Wang, and T.~Zhao (2014).
\newblock Sparse covariance matrix estimation with eigenvalue constraints.
\newblock {\em Journal of Computational and Graphical Statistics\/}~{\em 23\/}(2), 439--459.

\bibitem[\protect\citeauthoryear{Longford and Muth{\'e}n}{Longford and Muth{\'e}n}{1992}]{NiMu92-multilevel_factor_analysis}
Longford, N.~T. and B.~O. Muth{\'e}n (1992).
\newblock Factor analysis for clustered observations.
\newblock {\em Psychometrika\/}~{\em 57}, 581--597.

\bibitem[\protect\citeauthoryear{Lopes, Gamerman, and Salazar}{Lopes et~al.}{2011}]{Lo_et_al11-spatial_dynamic_factormodels}
Lopes, H.~F., D.~Gamerman, and E.~Salazar (2011).
\newblock Generalized spatial dynamic factor models.
\newblock {\em Computational Statistics \& Data Analysis\/}~{\em 55\/}(3), 1319--1330.

\bibitem[\protect\citeauthoryear{Lopuha{\"a}, Gares, and Ruiz-Gazen}{Lopuha{\"a} et~al.}{2023}]{Lo_et_al23-same_name_03}
Lopuha{\"a}, Hendrik, P., V.~Gares, and A.~Ruiz-Gazen (2023).
\newblock {S-estimation in Linear Models with Structured Covariance Matrices *}.
\newblock {\em {Annals of Statistics}\/}~{\em 51\/}(6), 2415--2439.

\bibitem[\protect\citeauthoryear{Lyons et~al.}{Lyons et~al.}{1988}]{Ly88-strongLLN_weaklycorrelated}
Lyons, R. et~al. (1988).
\newblock Strong laws of large numbers for weakly correlated random variables.
\newblock {\em Michigan Math. J\/}~{\em 35\/}(3), 353--359.

\bibitem[\protect\citeauthoryear{MacNab}{MacNab}{2011}]{Ma11-CARvariations}
MacNab, Y.~C. (2011).
\newblock On {G}aussian {M}arkov random fields and {B}ayesian disease mapping.
\newblock {\em Statistical Methods in Medical Research\/}~{\em 20\/}(1), 49--68.

\bibitem[\protect\citeauthoryear{Martini, Crossa, Toledo, and Cuevas}{Martini et~al.}{2020}]{Ma_et_al20-covmatHadamardProduct}
Martini, J.~W., J.~Crossa, F.~H. Toledo, and J.~Cuevas (2020).
\newblock On {H}adamard and {K}ronecker products in covariance structures for genotype$\times$ environment interaction.
\newblock {\em The Plant Genome\/}~{\em 13\/}(3), e20033.

\bibitem[\protect\citeauthoryear{Mayer and Zignago}{Mayer and Zignago}{2006}]{MaZi06-CEPIInotes}
Mayer, T. and S.~Zignago (2006).
\newblock Notes on {CEPII}’s distances measures.
\newblock electronic resource: https://mpra.ub.uni-muenchen.de/26469/1/MPRA{\_}paper{\_}26469.pdf.

\bibitem[\protect\citeauthoryear{Pourahmadi}{Pourahmadi}{1999}]{Po99-interpretable_unconstrained_parametrization}
Pourahmadi, M. (1999).
\newblock Joint mean-covariance models with applications to longitudinal data: Unconstrained parameterisation.
\newblock {\em Biometrika\/}~{\em 86\/}(3), 677--690.

\bibitem[\protect\citeauthoryear{Pourahmadi}{Pourahmadi}{2011}]{Po11-interpretable_unconstrained_parametrization02}
Pourahmadi, M. (2011).
\newblock {Covariance Estimation: The GLM and Regularization Perspectives}.
\newblock {\em Statistical Science\/}~{\em 26\/}(3), 369 -- 387.

\bibitem[\protect\citeauthoryear{Pourahmadi}{Pourahmadi}{2013}]{Po13-highDimCovmatEstims_review}
Pourahmadi, M. (2013).
\newblock {\em High-dimensional covariance estimation: with high-dimensional data}, Volume 882.
\newblock John Wiley \& Sons.

\bibitem[\protect\citeauthoryear{Raftery}{Raftery}{1995}]{Ra95-BIC_approx}
Raftery, A.~E. (1995).
\newblock Bayesian model selection in social research.
\newblock {\em Sociological methodology\/}, 111--163.

\bibitem[\protect\citeauthoryear{Rubin}{Rubin}{1976}]{Ru76-missing_at_random}
Rubin, D.~B. (1976).
\newblock Inference and missing data.
\newblock {\em Biometrika\/}~{\em 63\/}(3), 581--592.

\bibitem[\protect\citeauthoryear{Seaman, Galati, Jackson, and Carlin}{Seaman et~al.}{2013}]{Sh_et_al13-missingdata}
Seaman, S., J.~Galati, D.~Jackson, and J.~Carlin (2013).
\newblock {What Is Meant by “Missing at Random”?}
\newblock {\em Statistical Science\/}~{\em 28\/}(2), 257 -- 268.

\bibitem[\protect\citeauthoryear{Shanno}{Shanno}{1970}]{Sh70-BFGS04}
Shanno, D.~F. (1970).
\newblock Conditioning of quasi-{N}ewton methods for function minimization.
\newblock {\em Mathematics of Computation\/}~{\em 24\/}(111), 647--656.

\bibitem[\protect\citeauthoryear{Sun, Babu, and Palomar}{Sun et~al.}{2016}]{Su_et_al16-same_name_04}
Sun, Y., P.~Babu, and D.~P. Palomar (2016).
\newblock Robust estimation of structured covariance matrix for heavy-tailed elliptical distributions.
\newblock {\em IEEE Transactions on Signal Processing\/}~{\em 64\/}(14), 3576--3590.

\bibitem[\protect\citeauthoryear{Tastu, Pinson, and Madsen}{Tastu et~al.}{2013}]{Ta_et_al13-DCARspatiotemporal}
Tastu, J., P.~Pinson, and H.~Madsen (2013).
\newblock Space-time scenarios of wind power generation produced using a {G}aussian copula with parametrized precision matrix.

\bibitem[\protect\citeauthoryear{Thorson, Scheuerell, Shelton, See, Skaug, and Kristensen}{Thorson et~al.}{2015}]{Th_et_al15-spatial_factoranalysis}
Thorson, J.~T., M.~D. Scheuerell, A.~O. Shelton, K.~E. See, H.~J. Skaug, and K.~Kristensen (2015).
\newblock Spatial factor analysis: a new tool for estimating joint species distributions and correlations in species range.
\newblock {\em Methods in Ecology and Evolution\/}~{\em 6\/}(6), 627--637.

\bibitem[\protect\citeauthoryear{Tokuda, Goodrich, Van~Mechelen, Gelman, and Tuerlinckx}{Tokuda et~al.}{2011}]{To_et_al11-covmatvisualization}
Tokuda, T., B.~Goodrich, I.~Van~Mechelen, A.~Gelman, and F.~Tuerlinckx (2011).
\newblock Visualizing distributions of covariance matrices.
\newblock {\em Columbia Univ., New York, USA, Tech. Rep\/}, 18--18.

\bibitem[\protect\citeauthoryear{Turlach and Weingessel}{Turlach and Weingessel}{2019}]{BeWe19-quadprog}
Turlach, B.~A. and A.~Weingessel (2019).
\newblock {\em quadprog: Functions to Solve Quadratic Programming Problems}.
\newblock R package version 1.5-8.

\bibitem[\protect\citeauthoryear{Ver~Hoef, Hanks, and Hooten}{Ver~Hoef et~al.}{2018}]{Ve_et_al18-CARmodeltheory}
Ver~Hoef, J.~M., E.~M. Hanks, and M.~B. Hooten (2018).
\newblock On the relationship between conditional ({CAR}) and simultaneous ({SAR}) autoregressive models.
\newblock {\em Spatial statistics\/}~{\em 25}, 68--85.

\bibitem[\protect\citeauthoryear{Wall}{Wall}{2004}]{Wa04-CARexplanations}
Wall, M.~M. (2004).
\newblock A close look at the spatial structure implied by the {CAR} and {SAR} models.
\newblock {\em Journal of statistical planning and inference\/}~{\em 121\/}(2), 311--324.

\bibitem[\protect\citeauthoryear{Wang and Wall}{Wang and Wall}{2003}]{WaWa03-generalized_spatial_factormodels}
Wang, F. and M.~M. Wall (2003).
\newblock Generalized common spatial factor model.
\newblock {\em Biostatistics\/}~{\em 4\/}(4), 569--582.

\bibitem[\protect\citeauthoryear{Wei and Simko}{Wei and Simko}{2021}]{corrplot2021}
Wei, T. and V.~Simko (2021).
\newblock {\em R package 'corrplot': Visualization of a Correlation Matrix}.
\newblock (Version 0.92).

\bibitem[\protect\citeauthoryear{Zhou and Palomar}{Zhou and Palomar}{2024}]{ZhPa24-covFactorModel}
Zhou, R. and D.~P. Palomar (2024).
\newblock {\em covFactorModel: Covariance Matrix Estimation via Factor Models}.
\newblock R package version 0.1.0.

\end{thebibliography}


\begin{thebibliography}{}

\bibitem[\protect\citeauthoryear{Anderson and Burnham}{Anderson and Burnham}{2004}]{AnBu04-model_averaging}
Anderson, D. and K.~Burnham (2004).
\newblock {\em Model selection and multi-model inference}, Volume~63.
\newblock Springer New York.

\bibitem[\protect\citeauthoryear{Barv{\'\i}nek, Daler, and Francu}{Barv{\'\i}nek et~al.}{1991}]{Ba_et_al91-convergence_of_inverses}
Barv{\'\i}nek, E., I.~Daler, and J.~Francu (1991).
\newblock Convergence of sequences of inverse functions.
\newblock {\em Archivum Mathematicum\/}~{\em 27\/}(2), 201--204.

\bibitem[\protect\citeauthoryear{Chen}{Chen}{2023}]{Ch23-bootstrap_init}
Chen, Y.-C. (2023).
\newblock Statistical inference with local optima.
\newblock {\em Journal of the American Statistical Association\/}~{\em 118\/}(543), 1940--1952.

\bibitem[\protect\citeauthoryear{Erdős and Rényi}{Erdős and Rényi}{1959}]{ErRe1959randomgraphs}
Erdős, P. and A.~Rényi (1959).
\newblock On random graphs i.
\newblock {\em Publ. Math.\/}~{\em 6\/}(290-297), 18.

\bibitem[\protect\citeauthoryear{Folland}{Folland}{2005}]{fo05-multiTaylorExpansion}
Folland, G.~B. (2005).
\newblock Higher-order derivatives and {T}aylor’s formula in several variables.
\newblock {\em Preprint\/}, 1--4.
\newblock \url{https://sites.math.washington.edu/~folland/Math425/taylor2.pdf}.

\bibitem[\protect\citeauthoryear{Fosdick and Raftery}{Fosdick and Raftery}{2014}]{FoRa14-corlinearreg}
Fosdick, B.~K. and A.~E. Raftery (2014).
\newblock Regional probabilistic fertility forecasting by modeling between-country correlations.
\newblock {\em Demographic Research\/}~{\em 30\/}(35), 1011.

\bibitem[\protect\citeauthoryear{Goldfarb and Idnani}{Goldfarb and Idnani}{1983}]{GoId83-quadratic_optimization02}
Goldfarb, D. and A.~Idnani (1983).
\newblock A numerically stable dual method for solving strictly convex quadratic programs.
\newblock {\em Mathematical programming\/}~{\em 27\/}(1), 1--33.

\bibitem[\protect\citeauthoryear{Goldfarb and Idnani}{Goldfarb and Idnani}{2006}]{GoId06-quadratic_optimization01}
Goldfarb, D. and A.~Idnani (2006).
\newblock Dual and primal-dual methods for solving strictly convex quadratic programs.
\newblock In {\em Numerical Analysis: Proceedings of the Third IIMAS Workshop Held at Cocoyoc, Mexico, January 1981}, pp.\  226--239. Springer.

\bibitem[\protect\citeauthoryear{Hotelling}{Hotelling}{1953}]{Ha53-standard_corr_var_estim}
Hotelling, H. (1953).
\newblock New light on the correlation coefficient and its transforms.
\newblock {\em Journal of the Royal Statistical Society. Series B (Methodological)\/}~{\em 15\/}(2), 193--232.

\bibitem[\protect\citeauthoryear{Josse and Husson}{Josse and Husson}{2016}]{JoHu16-missingvalueimputation}
Josse, J. and F.~Husson (2016).
\newblock {missMDA}: A package for handling missing values in multivariate data analysis.
\newblock {\em Journal of Statistical Software\/}~{\em 70\/}(1), 1--31.

\bibitem[\protect\citeauthoryear{Ledoit and Wolf}{Ledoit and Wolf}{2003}]{LeWo03-CovMatCombination}
Ledoit, O. and M.~Wolf (2003).
\newblock Improved estimation of the covariance matrix of stock returns with an application to portfolio selection.
\newblock {\em Journal of empirical finance\/}~{\em 10\/}(5), 603--621.

\bibitem[\protect\citeauthoryear{Mardia and Marshall}{Mardia and Marshall}{1984}]{MaMa84-1sampleMLE}
Mardia, K.~V. and R.~J. Marshall (1984).
\newblock Maximum likelihood estimation of models for residual covariance in spatial regression.
\newblock {\em Biometrika\/}~{\em 71\/}(1), 135--146.

\bibitem[\protect\citeauthoryear{Munkres}{Munkres}{2000}]{Mu20_topology_uniform_extension}
Munkres, J. (2000).
\newblock {\em Topology}.
\newblock Featured Titles for Topology. Prentice Hall, Incorporated.

\bibitem[\protect\citeauthoryear{Petersen and Pedersen}{Petersen and Pedersen}{2008}]{PeP08}
Petersen, K.~B. and M.~S. Pedersen (2008).
\newblock The {M}atrix {C}ookbook.
\newblock Technical report, Technical University of Denmark.

\bibitem[\protect\citeauthoryear{Rockafellar and Wets}{Rockafellar and Wets}{2009}]{Rockafellar&2009}
Rockafellar, R. and R.-B. Wets (2009).
\newblock {\em Variational Analysis\/} (1st ed.).
\newblock Springer.

\bibitem[\protect\citeauthoryear{Rubin}{Rubin}{1976}]{Ru76-missing_at_random}
Rubin, D.~B. (1976).
\newblock Inference and missing data.
\newblock {\em Biometrika\/}~{\em 63\/}(3), 581--592.

\bibitem[\protect\citeauthoryear{Seabrook and Wiskott}{Seabrook and Wiskott}{2023}]{SeWi23-spectraltheory_MarkovChain}
Seabrook, E. and L.~Wiskott (2023).
\newblock A tutorial on the spectral theory of markov chains.
\newblock {\em Neural Computation\/}~{\em 35\/}(11), 1713--1796.

\bibitem[\protect\citeauthoryear{Seaman, Galati, Jackson, and Carlin}{Seaman et~al.}{2013}]{Sh_et_al13-missingdata}
Seaman, S., J.~Galati, D.~Jackson, and J.~Carlin (2013).
\newblock {What Is Meant by “Missing at Random”?}
\newblock {\em Statistical Science\/}~{\em 28\/}(2), 257 -- 268.

\bibitem[\protect\citeauthoryear{Turlach and Weingessel}{Turlach and Weingessel}{2019}]{BeWe19-quadprog}
Turlach, B.~A. and A.~Weingessel (2019).
\newblock {\em quadprog: Functions to Solve Quadratic Programming Problems}.
\newblock R package version 1.5-8.

\end{thebibliography}

\end{document}

\begin{frontmatter}
\title{Supplementary Material for\\\\A Structured Estimator for large Covariance Matrices in the Presence of Pairwise and Spatial Covariates}
\runtitle{A Structured Estimator for large Covariance Matrices}

\begin{aug}
\author[A]{\fnms{Martin}~\snm{Metodiev}\ead[label=e1]{martin.metodiev@doctorant.uca.fr}},
\author[B]{\fnms{Marie}~\snm{Perrot-Dock\`{e}s}\ead[label=e2]{marie.perrot-dockees@u-paris.fr}},
\author[H]{\fnms{Sarah}~\snm{Ouadah}\ead[label=e3]{sarah.ouadah@sorbonne-universite.fr}},
\author[G]{\fnms{Bailey K.}~\snm{Fosdick}\ead[label=e4]{bailey.fosdick@cuanschutz.edu}},
\author[H]{\fnms{Stéphane}~\snm{Robin}\ead[label=e5]{stephane.robin@sorbonne-universite.fr}},
\author[A,F]{\fnms{Pierre}~\snm{Latouche}\ead[label=e6]{Pierre.LATOUCHE@uca.fr}} \and
\author[D]{\fnms{Adrian E.}~\snm{Raftery}\ead[label=e7]{raftery@uw.edu}}
%%%%%%%%%%%%%%%%%%%%%%%%%%%%%%%%%%%%%%%%%%%%%%
%% Addresses                                %%
%%%%%%%%%%%%%%%%%%%%%%%%%%%%%%%%%%%%%%%%%%%%%%
\address[A]{Université Clermont Auvergne, Laboratoire de Mathématiques Blaise Pascal,\\\printead[presep={ }]{e1,e6}}
%\address[E]{Corresponding author\printead[presep={,\ }]{e1}}
\address[B]{Université Paris Cité, CNRS, MAP5, F-75006 Paris, France\printead[presep={,\ }]{e2}}
\address[H]{Sorbonne Université, Laboratoire de Probabilités, Statistique et Modélisation (LPSM)\printead[presep={,\ }]{e3,e5}}
\address[G]{GTI Energy\printead[presep={,\ }]{e4}}
\address[F]{Institut universitaire de France (IUF)\printead[presep={,\ }]{e6}}
\address[D]{University of Washington, Department of Statistics\printead[presep={,\ }]{e7}}
\runauthor{M. Metodiev et al.}
\end{aug}

\end{frontmatter}

\section*{Appendix A: derivatives of the likelihood}

We are going to calculate the partial derivatives of of the likelihood in the case that the spatial effect is given by the CAR model, meaning that \begin{align*}
    R(\alpha&,\beta,\delta)=\delta_D Q_{\beta_D}^{-1}\left(I_d-\beta_D M_1\right)^{-1}M_2^{-1}Q^{-1}_{\beta_D}+\alpha_AF_A+\alpha_BF_B+\alpha_CF_C+\alpha_EI_d,\\\alpha_A&,\alpha_B,\alpha_C,\delta_D,\alpha_E>0,\quad \alpha_A+\alpha_B+\alpha_C+\delta_D+\alpha_E=1,
\end{align*}  where $(M_1)_{i,j}$ is equal to the reciprocal of the number of neighbors of country $i$ if country $i$ and $j$ are neighbors. Moreover, $M_2$ denotes a diagonal matrix including the number of neighbors of each country, such that \begin{align*}
    (M_2)_{i,j}=\begin{cases}
        \sum_{e=1}^d M_{i,e}&i=j\\0&i\neq j
    \end{cases},\; (M_1)_{i,j}=\frac{M_{i,j}}{(M_2)_{i,j}}.
\end{align*} $Q_{\beta_D}$ is a non-negative diagonal matrix chosen such that all diagonal entries of \\$\Gamma(\beta_D)^{-1}=Q_{\beta_D}^{-1}(I_d-\beta_D M_1)^{-1}M_2^{-1}Q_{\beta_D}^{-1}$ are equal to 1.

\paragraph*{\textbf{Computing an explicit solution for $\Gamma(\beta_D)^{-1}$}}

\iffalse
Some results, which will be used:
\begin{itemize}
    \item[(I)] Let $M_1$ be a stochastic matrix of a reversible and finite Markov chain. Let $\lambda_i,1\leq i\leq d$ denote the eigenvalues of $M_1$. All of the eigenvalues of $M_1$ are real numbers and $\max_i|\lambda_i|\leq 1$ by the Perron Frobenius Theorem \citep[Theorem 7]{SeWi23-spectraltheory_MarkovChain}. In fact, $M_1$ only has real eigenvalues \citep[Theorem 14]{SeWi23-spectraltheory_MarkovChain}, so it is even diagonalizeable.

    \item[(II)] For any $M_1$, if $\max_i |\lambda_i|<1$ holds for the eigenvalues of $M_1$, we have \citep[Section 3.4]{PeP08}\begin{align}
        (I_d-M_1)^{-1}=\sum^\infty_{k=0}M_1^k.
    \end{align}
\end{itemize}
\fi

All rows of $M_1$ are normalized such that their sum is equal to 1, so $M_1$ is a stochastic matrix. The condition $(M_2)_{i,i} (M_1)_{i,j}=(M_2)_{j,j} (M_1)_{j,i}=1$ is fulfilled by default, so $M_1$ is the stochastic matrix of a reversible, finite Markov chain. Thus, by the Perron Frobenius Theorem \citep[Theorem 7]{SeWi23-spectraltheory_MarkovChain}, all of the eigenvalues of $M_1$ are real numbers and are smaller or equal to one. As $\beta_D <1$, this directly implies that the absolute values of the eigenvalues of $\beta_D M_1$ are all strictly smaller than 1. Thus, using \citet[Section 3.4]{PeP08} we get  \begin{align}
    (I_d-\beta_D M_1)^{-1}=\sum^\infty_{k=0}\beta^k_D M_1^k.
\end{align}

$M_1$ is the stochastic matrix of a reversible, finite Markov chain, thus  $M_1$ is diagonalizeable \citep[Theorem 14]{SeWi23-spectraltheory_MarkovChain}. Hence, there exists a matrix $U$ such that\begin{align}
    \beta_D M_1=U \text{diag}(\beta_D \lambda_1,\dots,\beta_D \lambda_d) U^{-1}.
\end{align}

In total \begin{align*}
    (I_d-\beta_D M_1)^{-1}M_2^{-1}&=(\sum^\infty_{k=0}\beta^k_DM_1^k)M_2^{-1}=(I_d+\sum^\infty_{k=1}\beta^k_DM_1^k)M_2^{-1}\\&=(I_d+\sum^\infty_{k=1}U \text{diag}(\beta^k_D\lambda_1^k,\dots,\beta^k_D\lambda_d^k)U^{-1})M_2^{-1}\\&=(I_d+U \text{diag}(\sum^\infty_{k=1}\beta^k_D\lambda_1^k,\dots,\sum^\infty_{k=1}\beta^k_D\lambda_d^k)U^{-1})M_2^{-1}\\&=(I_d+U \text{diag}(-1+\sum^\infty_{k=0}\beta^k_D\lambda_1^k,\dots,-1+\sum^\infty_{k=0}\beta^k_D\lambda_d^k)U^{-1})M_2^{-1}\\&=(I_d+U\text{diag}(\frac{\beta_D\lambda_1}{1-\beta_D\lambda_1},\dots,\frac{\beta_D\lambda_d}{1-\beta_D\lambda_d})U^{-1})M_2^{-1}.
\end{align*}

This has elements \begin{align*}
    ((I_d-\beta_D M_1)^{-1}M_2^{-1})_{i,j}=\frac{1}{v_j}(1_{i,j}+\sum^d_{k=1}\frac{\beta_D \lambda_k}{1-\beta_D\lambda_k} U_{i,k}U_{k,j}^{-1}).
\end{align*} Here, $1_{i,j}$ denotes the Kronecker-delta and $v_j$ denotes the number of neighbors of country $j$. The correlation matrix is thus defined by\begin{align*}
    (\Gamma(\beta_D)^{-1})_{i,j}=\sqrt{\frac{v_i}{v_j}}\frac{\sum^d_{k=1}\frac{\beta_D \lambda_k}{1-\beta_D\lambda_k} U_{i,k}U_{k,j}^{-1}}{\sqrt{1+\sum^d_{k=1}\frac{\beta_D \lambda_k}{1-\beta_D\lambda_k} U_{i,k}U_{k,i}^{-1}}\sqrt{1+\sum^d_{k=1}\frac{\beta_D \lambda_k}{1-\beta_D\lambda_k} U_{j,k}U_{k,j}^{-1}}}.
\end{align*}  Note that this expression decreases computation time, since the only matrix that needs to be inversed to compute it is $U$, which does not depend on $\beta_D$, and is a direct result of a computationally inexpensive diagonalization procedure.

\begin{remark}
The results we obtained can be used to show that $\Gamma$ is uniformly continuous: suppose without loss of generality that the graph is connected, meaning that $M_1$ is irreducible (if not, we can look at the correlation matrix of the individual connected components). 
If $\beta_D=0$, the correlation between node $i$ and $j$ is always 0. The entries of $U$ are the normalized eigenvectors of $M_1$. 1 is an eigenvalue of $M_1$ with eigenvector $(1,\dots, 1)$, so $\beta_D\uparrow 1$ gives a correlation of \begin{align}
    \sqrt{\frac{v_i}{v_j}}\sqrt{\frac{U_{1,j}^{-1}}{U_{1,i}^{-1}}}=\begin{cases}1&\exists\text{ path from }i\text{ to }j,\\0&\text{else,}\end{cases},
\end{align} since $U_{1\cdot}$ is proportional to $(v_1,\dots,v_d)$, the left-eigenvector of $M_1$ \citep[notice that there only exists one eigenvector to the eigenvalue 1 in an irreducible, finite Markov chain; for a proof, see for instance][Theorem 6]{SeWi23-spectraltheory_MarkovChain}. The covariances and variances, on the other hand, go to infinity, so they are hard to interpret.

The domain of $\Gamma$ is $(0,1)$ and we have just shown that the limits $\lim_{\beta_D\uparrow1}\Gamma(\beta_D)$ and $\lim_{\beta_D\downarrow0}\Gamma(\beta_D)$ are well defined, so $\Gamma$ is uniformly continuous, which is an assumption needed for Theorem 3.2.
\end{remark}

{ We are going to calculate the derivatives in two scenarios: in the first, we consider that the variance and mean parameters are given by known estimators and we are only maximizing the likelihood with respect to $(\alpha,\beta,\delta)$. In the second, we are trying to maximize the likelihood with respect to the correlation, mean and variance parameters $(\mu,\sigma,\alpha,\beta,\delta)$.}

\paragraph*{\textbf{Calculating the derivatives of the likelihood when $\hat{\mu}$ and $\hat{\sigma}$ are given}}

It turns out that the likelihood (or a function proportional to it) can be expressed as a function of the estimated standardized errors \begin{align}
    \hat{\varepsilon}_t=\text{diag}(\hat{\sigma}_t)^{-1}(Y_t-\hat{\mu}_t).
\end{align} We simplify {\begin{align*}
    L&(\alpha,\beta,\delta)=\prod^{T}_{t=1}\text{MVN}_d(Y_t;\hat{\mu_t},\text{diag}(\hat{\sigma_t})R(\alpha,\beta,\delta)\text{diag}(\hat{\sigma_t}))\\&\propto\prod^{T}_{t=1}\frac{\exp\left(-\frac{1}{2}(Y_t-\hat{\mu}_t)^\intercal(\text{diag}(\hat{\sigma_t})R(\alpha,\beta,\delta)\text{diag}(\hat{\sigma_t}))^{-1}(Y_t-\hat{\mu}_t)\right)}{\sqrt{|\text{diag}(\hat{\sigma_t})R(\alpha,\beta,\delta)\text{diag}(\hat{\sigma_t})|}}\\&=\prod^{T}_{t=1}\frac{1/|\text{diag}(\hat{\sigma_t})|^2}{\sqrt{|R(\alpha,\beta,\delta)|}}\exp(-\frac{1}{2}(\text{diag}(\hat{\sigma_t})^{-1}(Y_t-\hat{\mu}_t))^\intercal R(\alpha,\beta,\delta)^{-1}(\text{diag}(\hat{\sigma_t})^{-1}(Y_t-\hat{\mu}_t)))\\&\propto\prod^{T}_{t=1}\frac{1}{\sqrt{|R(\alpha,\beta,\delta)|}}\exp\left(-\frac{1}{2}\hat{\varepsilon}^\intercal_tR(\alpha,\beta,\delta)^{-1}\hat{\varepsilon}_t\right),\end{align*}}

where $|R(\alpha,\beta,\delta)|$ denotes the determinant of $R(\alpha,\beta,\delta)$. Notice that this is always the case, independent from the number of components, the definition of $\Gamma$, and the size of $\delta$.

Let $l(\alpha,\beta,\delta)$ denote $T$ times the logarithm of the square of the above function,\begin{align}
    l(\alpha,\beta,\delta):=\frac{1}{T}\log\prod^{T}_{t=1}\frac{1}{|R(\alpha,\beta,\delta)|}\exp\left(-\hat{\varepsilon}^\intercal_tR(\alpha,\beta,\delta)^{-1}\hat{\varepsilon}_t\right).
\end{align}

Going back to our particular case, all parameters sum to one, so $\alpha_E=1-\alpha_A-\alpha_B-\alpha_C-\delta_D$. We thus only need the derivatives of $l$ for the latter parameters. Derivatives were calculated using the matrix cookbook \citep{PeP08}. The notation $R(\alpha,\beta,\delta)$ was shortened to $R$ to improve readability.

%\newpage

\paragraph*{\textbf{List of derivatives in the case that $\hat{\mu},\hat{\sigma}$ are given}} \begin{align*}\Gamma^{-1}&=\delta_D Q_{\beta_D}^{-1}(I_d+UD_{\beta_D} U^{-1})M_2^{-1}Q_{\beta_D}^{-1}\\R&=\Gamma^{-1}+(1-\delta_D-\alpha_A-\alpha_B-\alpha_C)I_d+\alpha_AF_A+\alpha_BF_B+\alpha_CF_C\\S_T&=\frac{1}{T}\sum^T_{t=1}\hat{\varepsilon}_t\hat{\varepsilon}_t^\intercal\\ l(\alpha,\beta,\delta)&=-\log(|R|)-\text{tr}(S_TR^{-1})\\
    \frac{\partial}{\partial\alpha_A}l(\alpha,\beta,\delta)&= -\text{tr}((F_A-I_d)R^{-1})+\text{tr}(S_TR^{-1} (F_A-I_d)R^{-1})\quad(\text{analogous for }\alpha_B,\alpha_C)\\
    \frac{\partial}{\partial\delta_D}l(\alpha,\beta,\delta)&= -\text{tr}((Q_{\beta_D}^{-1}(I_d+UD_{\beta_D} U^{-1})M_2^{-1}Q_{\beta_D}^{-1}-I_d)R^{-1})\\&+\text{tr}(S_TR^{-1} (Q_{\beta_D}^{-1}(I_d+UD_{\beta_D} U^{-1})M_2^{-1}Q_{\beta_D}^{-1}-I_d)R^{-1})\\
    (D_{\beta_D})_{i,i}&=\frac{\beta_D\lambda_i}{1-\beta_D\lambda_i}\\
    (Q^{-1}_{\beta_D})_{i,i}&=\frac{1}{\sqrt{((I_d+UD_{\beta_D} U^{-1})M_2^{-1})_{i,i}}}\\
    (\frac{\partial}{\partial\beta_D}D_{\beta_D})_{i,i}&=\frac{\lambda_i}{(1-\beta_D\lambda_i)^2}\\
    (\frac{\partial}{\partial\beta_D}Q^{-1}_{\beta_D})_{i,i}&=-\frac{1}{2}\left(\frac{1}{\sqrt{((I_d+UD_{\beta_D} U^{-1})M_2^{-1})_{i,i}}}\right)^3(U(\frac{\partial}{\partial\beta}D_{\beta_D} )U^{-1}M_2^{-1})_{i,i}\\
    \frac{\partial}{\partial\beta_D}(\Gamma(\beta_D)^{-1})&=(\frac{\partial}{\partial\beta_D}Q^{-1}_{\beta_D}) (I+UD_{\beta_D} U^{-1})M_2^{-1}Q_{\beta_D}^{-1}\\&+Q^{-1}_{\beta_D} U(\frac{\partial}{\partial\beta_D} D_{\beta_D}) U^{-1}M_2^{-1}Q_{\beta_D}^{-1}\\&+Q^{-1}_{\beta_D} (I_d+UD_{\beta_D} U^{-1})M_2^{-1}(\frac{\partial}{\partial\beta_D}Q_{\beta_D}^{-1})\\\frac{\partial}{\partial\beta_D}l(\alpha,\beta,\delta)&=-\text{tr}(\delta_D\frac{\partial}{\partial\beta_D}(\Gamma(\beta_D)^{-1})R^{-1})\\&+\text{tr}(S_TR^{-1} \delta_D\frac{\partial}{\partial\beta_D}(\Gamma(\beta_D)^{-1})R^{-1})
\end{align*}
\newpage
\paragraph*{\textbf{Calculating the derivatives of the likelihood when the means and variances are unknown}}

The calculations are very similar, though the likelihood does not simplify as much. We thus only need to calculate the derivatives of \begin{align*}
    l_2(\alpha,\beta,\delta,\mu,\sigma):&=\frac{1}{T}\log\prod^{T}_{t=1}\frac{1/|\text{diag}(\sigma_t)|}{|R(\alpha,\beta,\delta)|}\exp\left(-\varepsilon^\intercal_tR(\alpha,\beta,\delta)^{-1}\varepsilon_t\right)\\&=l(\alpha,\beta,\delta)-\frac{1}{T}\sum^T_{t=1}\log(|\text{diag}(\sigma_t)|).
\end{align*} Derivatives with respect to $(\alpha,\beta,\delta)$ are thus identical to the derivatives that we already calculated. Derivatives with respect to $\mu_t$ are given by \begin{align*}
    \frac{\partial}{\partial\mu_t}l_2(\alpha,\beta,\delta,\mu,\sigma)=\frac{1}{T}(\text{diag}(\sigma_t)^{-1}\mu_t)^\intercal R(\alpha,\beta,\delta)^{-1}\varepsilon_t+\frac{1}{T}\varepsilon_t^\intercal R(\alpha,\beta,\delta)^{-1}\text{diag}(\sigma_t)^{-1}\mu_t.
\end{align*}

Derivatives with respect to $\sigma_t$ are given by\begin{align*}
    \frac{\partial}{\partial\sigma_t}l_2(\alpha,\beta,\delta,\mu,\sigma)&=\frac{1}{T}(\text{diag}(\sigma_t^2)^{-1}(Y_t-\mu_t))^\intercal R(\alpha,\beta,\delta)^{-1}\varepsilon_t\\&+\frac{1}{T}\varepsilon_t^\intercal R(\alpha,\beta,\delta)^{-1}\text{diag}(\sigma_t^2)^{-1}(Y_t-\mu_t)-\frac{1}{T}\sum^T_{t=1}\text{tr}(\text{diag}(\sigma_t)^{-1}).
\end{align*}

Often, $\mu_t$ and $\sigma_t$ are not estimated directly, since there are some underlying conditions that simplify the estimation procedure. In an autoregressive model, for example, $\mu_t$ and $\sigma_t$ can be expressed by their predecessors, so we are really only optimizing over 3 autoregressive parameters. The respective changes to the derivatives can be added via the multidimensional chain rule. 

\section*{Appendix B: proofs}

\begin{proof}[Proof of Theorem 3.2]
    The model is identifiable if, and only if any two pairs of parameters $(\alpha,\beta,\delta)\neq(\alpha',\beta',\delta')$ give two different correlation matrices. This is the case if, and only if it is not the case that\begin{align}
        \delta\Gamma(\beta)^{-1}+\sum^K_{k=0}\alpha_k F_k= \delta'\Gamma(\beta')^{-1}+\sum^K_{k=0}\alpha_k' F_k
    \end{align} for any $(\alpha,\beta,\delta)\neq(\alpha',\beta',\delta')$. If there exists such $(\alpha,\beta,\delta),(\alpha',\beta',\delta')$, then there are two possibilities: either $\beta=\beta'$, in which case \begin{align}
        (\delta-\delta')\Gamma(\beta)^{-1}+\sum^K_{k=0}(\alpha_k-\alpha_k')F_k=0,
    \end{align} implying dependence between the matrices, or $\beta\neq\beta'$, in which case the LP  \begin{equation*}
\begin{array}{ll@{}ll}
\mathrm{max}_{\alpha,\alpha',\delta,\delta'}  & \delta+\delta'\\
\mathrm{such}\; \mathrm{that } &\displaystyle\sum\limits_{k=0}^{K}(\alpha_k-\alpha_k')F_k+\delta\Gamma(\beta)^{-1}    = \delta'\Gamma(\beta')^{-1}\\&\displaystyle\sum\limits_{k=0}^{K}\alpha_k+\delta=\displaystyle\sum\limits_{k=0}^{K}\alpha_k'+\delta'=1,\quad\alpha_0,\dots,\alpha_K,\alpha_0',\dots,\alpha_K',\delta,\delta'\geq0
\end{array}
\end{equation*}  has output strictly larger than 0 due to $(\alpha,\beta,\delta),(\alpha',\beta',\delta')$ being a feasible point.
\end{proof}

\begin{proof}[Proof of Theorem 3.4]

\textbf{Part 1: strong consistency}

We are going to consider the function \begin{align}\label{eq:apprx_loglik}
    l(\alpha,\beta,\delta):&=\frac{1}{T}\log\prod^{T}_{t=1}\frac{1}{|R(\alpha,\beta,\delta)|}\exp\left(-\hat{\varepsilon}^\intercal_tR(\alpha,\beta,\delta)^{-1}\hat{\varepsilon}_t\right)\nonumber\\&=-\log(|R(\alpha,\beta,\delta)|)-\text{tr}\left(\left(\frac{1}{T}\sum^{T}_{t=1}\hat{\varepsilon}_t\hat{\varepsilon}_t^\intercal\right)R(\alpha,\beta,\delta)^{-1}\right),
\end{align} which we are trying to maximize. Notice that Equation \eqref{eq:apprx_loglik} is not the log-likelihood of the model, because the variance and mean parameters are estimated. However, the usual proofs for the MLE don't change, which is what we are going to show.

First, we are going to show that the MLE $(\hat{\alpha}_{\text{MLE}},\hat{\beta}_{\text{MLE}},\hat{\delta}_{\text{MLE}})$ is strongly consistent. 
Let $(\alpha^{\star},\beta^{\star},\delta^{\star})$ be the true model parameter. By Condition (C1) $S_T:=\frac{1}{T}\sum^{T}_{t=1}\hat{\varepsilon}_t\hat{\varepsilon}_t^\intercal$ converges to $R(\alpha^{\star},\beta^{\star},\delta^{\star})$ almost surely. From now on, we are only considering sample values from the set with probability 1 for which this convergence takes place.

$(\hat{\alpha}_{\text{MLE}},\hat{\beta}_{\text{MLE}},\hat{\delta}_{\text{MLE}})$ cannot diverge to a point for which $|R(\hat{\alpha}_{\text{MLE}},\hat{\beta}_{\text{MLE}},\hat{\delta}_{\text{MLE}})|\stackrel{T\to\infty}\to0$. If that were the case,\begin{align*}
    \lim_{T\to\infty}l(\hat{\alpha}_{\text{MLE}},\hat{\beta}_{\text{MLE}},\hat{\delta}_{\text{MLE}})=\lim_{T\to\infty}(&-\log(|R(\hat{\alpha}_{\text{MLE}},\hat{\beta}_{\text{MLE}},\hat{\delta}_{\text{MLE}})|)\\&-\text{tr}(S_TR(\hat{\alpha}_{\text{MLE}},\hat{\beta}_{\text{MLE}},\hat{\delta}_{\text{MLE}})^{-1}))\\=\lim_{T\to\infty}(&-\log(|R(\hat{\alpha}_{\text{MLE}},\hat{\beta}_{\text{MLE}},\hat{\delta}_{\text{MLE}})|)\\&-\text{tr}(R(\alpha^{\star},\beta^{\star},\delta^{\star})R(\hat{\alpha}_{\text{MLE}},\hat{\beta}_{\text{MLE}},\hat{\delta}_{\text{MLE}})^{-1}))=-\infty,
\end{align*} since the latter term is $"$faster$"$ than the former. 

Thus, there exists a real number $c>0$ such that the set $C:=\{(\alpha,\beta,\delta):|R(\alpha,\beta,\delta)|\geq c\}$ contains $(\hat{\alpha}_{\text{MLE}},\hat{\beta}_{\text{MLE}},\hat{\delta}_{\text{MLE}})$ for every $T$. We have uniform convergence of $l$ to the function \begin{align}\label{eq:apprx_loglik_limit}
    f(\alpha,\beta,\delta):=-\log(|R(\alpha,\beta,\delta)|)-\text{tr}(R(\alpha^{\star},\beta^{\star},\delta^{\star})R(\alpha,\beta,\delta)^{-1})
\end{align}within $C$, as\begin{align*}
    \|l(\alpha,\beta,\delta)&-f(\alpha,\beta,\delta)\|_{C,\infty}=\|\text{tr}((S_T-R(\alpha^{\star},\beta^{\star},\delta^{\star}))R(\alpha,\beta,\delta)^{-1})\|_{C,\infty}\\&\leq \sqrt{|\text{tr}((S_T-R(\alpha^{\star},\beta^{\star},\delta^{\star}))^2)|\cdot\|\text{tr}(R(\alpha,\beta,\delta)^{-2})\|_{C,\infty}}\stackrel{T\to\infty}\to0,
\end{align*} where $\|\cdot\|_{C,\infty}$ denotes the supremum-norm on $C$. We recognize the shape of the logarithmic density function of an (improper) inverse Wishart-distribution with parameters $\text{InvWis}(-d,R(\alpha^{\star},\beta^{\star},\delta^{\star}))$ and unique mode $R(\alpha^{\star},\beta^{\star},\delta^{\star})$ in Expression \eqref{eq:apprx_loglik_limit}. Thus, $(\alpha^\star,\beta^\star,\delta^\star)$ is the unique mode of $f$ (uniqueness follows from Condition (A1), identifiability). 

Since $l$ and $f$ are uniformly continuous on $C$ (follows from Condition B1) they have a unique uniformly continuous extension to the closure of $C$ (a variation of this statement is given as an exercise in \citet[Chapter 7, Exercise 2]{Mu20_topology_uniform_extension}). Since their domain is also bounded, their unique uniformly continuous extensions have a compact domain. By continuity and uniform convergence we have epi-convergence of $-l,-f$ \citep[Proposition 7.15]{Rockafellar&2009}. Epi-convergence in a compact domain implies that the argmin of $-l$, which is the MLE, converges to the argmin of $-f$, which is $(\alpha^\star,\beta^\star,\delta^\star)$ \citep[this is a consequence of][Theorem 7.33]{Rockafellar&2009}. In particular, $R(\hat{\alpha}_{\text{MLE}},\hat{\beta}_{\text{MLE}},\hat{\delta}_{\text{MLE}})$ converges to $R(\alpha^\star,\beta^\star,\delta^\star)$. This implies convergence of $(\hat{\alpha}_{\text{SCE}},\hat{\beta}_{\text{SCE}},\hat{\delta}_{\text{SCE}})$ by continuity and uniform convergence of $l$ on all compact subsets of its domain \citep[this is a consequence of Condition (A1), which implies injectivity of $l$ and $f$, and][Corollary 1]{Ba_et_al91-convergence_of_inverses}. Since all of the calculations were made on a set with probability 1, it follows that $R(\alpha^{(0)},\beta^{(0)},\delta^{(0)})$ converges to the true correlation matrix as well, and by definition\begin{align*}
    l(\alpha^{(0)},\beta^{(0)},\delta^{(0)})\leq l(\hat{\alpha}_{\text{SCE}},\hat{\beta}_{\text{SCE}},\hat{\delta}_{\text{SCE}})\leq l(\hat{\alpha}_{\text{MLE}},\hat{\beta}_{\text{MLE}},\hat{\delta}_{\text{MLE}}),
\end{align*} so \begin{align*}
    l(\hat{\alpha}_{\text{SCE}},\hat{\beta}_{\text{SCE}},\hat{\delta}_{\text{SCE}})\stackrel{t\to\infty}\to f(\alpha^{\star},\beta^{\star},\delta^{\star}).
\end{align*} $(\hat{\alpha}_{\text{SCE}},\hat{\beta}_{\text{SCE}},\hat{\delta}_{\text{SCE}})$ is a strongly consistent estimator of $(\alpha^{\star},\beta^{\star},\delta^{\star})$, and thus we have (by Condition (B1) and the continuous mapping theorem) that $R(\hat{\alpha}_{\text{SCE}},\hat{\beta}_{\text{SCE}},\hat{\delta}_{\text{SCE}})$ is a strongly consistent estimator of $R(\alpha^{\star},\beta^{\star},\delta^{\star})$.

\textbf{Part 2: asymptotic normality}

We calculate a first order Taylor series expansion of the gradient of $l$ around $(\alpha^{\star},\beta^{\star},\delta^{\star})$. By applying \cite[Corollary 1]{fo05-multiTaylorExpansion} $d$ times, we get that $\triangledown l(\hat{\alpha}_{\text{SCE}},\hat{\beta}_{\text{SCE}},\hat{\delta}_{\text{SCE}})=$ \begin{align*}
    \triangledown l(\alpha^{\star},\beta^{\star},\delta^{\star})+H_l(\alpha^{\star},\beta^{\star},\delta^{\star})\cdot((\alpha^{\star},\beta^{\star},\delta^{\star})-(\hat{\alpha}_{\text{SCE}},\hat{\beta}_{\text{SCE}},\hat{\delta}_{\text{SCE}}))+R_T,
\end{align*} with $H_l$ the Hessian of $l$ and \begin{align*}
    \sum^d_{k=1}|(R_T)_i|\leq \sum^d_{i=1}\left(\max_W\{|\triangledown l(\alpha,\beta,\delta)_i|\}\cdot\sum_{k=1}^{K+G+1}|(\alpha^{\star},\beta^{\star},\delta^{\star})_k-(\hat{\alpha}_{\text{SCE}},\hat{\beta}_{\text{SCE}},\hat{\delta}_{\text{SCE}})_k|\right),
\end{align*} where \begin{align*}
    W:=\Big\{(\alpha,\beta,\delta)|&\sum_{k=1}^{K+G+1}|(\alpha^{\star},\beta^{\star},\delta^{\star})_k-(\alpha,\beta,\delta)_k|\leq \\&\sum_{k=1}^{K+G+1}|(\alpha^{\star},\beta^{\star},\delta^{\star})_k-(\hat{\alpha}_{\text{SCE}},\hat{\beta}_{\text{SCE}},\hat{\delta}_{\text{SCE}})_k|\Big\}.
\end{align*}

Using the fact that $\triangledown l(\hat{\alpha}_{\text{SCE}},\hat{\beta}_{\text{SCE}},\hat{\delta}_{\text{SCE}})=0$ (Condition A2) and rearranging and solving for $(\hat{\alpha}_{\text{SCE}},\hat{\beta}_{\text{SCE}},\hat{\delta}_{\text{SCE}})$ gives\begin{align}\label{eq:Hessian_bound}
    \sqrt{T}((\hat{\alpha}_{\text{SCE}},\hat{\beta}_{\text{SCE}},\hat{\delta}_{\text{SCE}})-(\alpha^{\star},\beta^{\star},\delta^{\star}))=-\sqrt{T}H_l(\alpha^{\star},\beta^{\star},\delta^{\star})^{-1}\cdot(\triangledown l(\alpha^{\star},\beta^{\star},\delta^{\star})+R_T).
\end{align}

The expression $-H_l(\alpha^{\star},\beta^{\star},\delta^{\star})^{-1}$ converges almost surely, since\begin{align*}
    (\frac{\partial^2}{\partial\alpha_k\partial\alpha_j}l)(\alpha_{\text{TRUE}},\beta_{\text{TRUE}}&,\delta_{\text{TRUE}})=-(\frac{\partial^2}{\partial\alpha_k\partial\alpha_j}\log|R|)(\alpha^{\star},\beta^{\star},\delta^{\star})\\&-\text{tr}(S_T(\frac{\partial^2}{\partial\alpha_k\partial\alpha_j}R^{-1})(\alpha^{\star},\beta^{\star},\delta^{\star}))\\\stackrel{T\to\infty}\to\nonumber&-(\frac{\partial^2}{\partial\alpha_k\partial\alpha_j}\log|R|)(\alpha^{\star},\beta^{\star},\delta^{\star})\\&-\text{tr}(R(\alpha^{\star},\beta^{\star},\delta^{\star})(\frac{\partial^2}{\partial\alpha_k\partial\alpha_j}R^{-1})(\alpha^{\star},\beta^{\star},\delta^{\star}))\\&=-2\cdot I(\alpha^{\star},\beta^{\star},\delta^{\star})_{k,j},
\end{align*} and the same goes for partial derivatives with respect to $\beta$ (due to Condition B2) and $\delta$. 

Similarly, convergence in distribution of $\sqrt{T}\triangledown l(\alpha^{\star},\beta^{\star},\delta^{\star})$ follows from applying Condition (C2) to {
\begin{align*}
    &\sqrt{T}(\frac{\partial}{\partial\alpha_k}l)(\alpha^{\star},\beta^{\star},\delta^{\star})\\&=\sqrt{T}\Big(-(\frac{\partial}{\partial\alpha_k}\log|R|)(\alpha^{\star},\beta^{\star},\delta^{\star})-\text{tr}(S_T(\frac{\partial}{\partial\alpha_k}R^{-1})(\alpha^{\star},\beta^{\star},\delta^{\star}))\Big)\nonumber\\&=\sqrt{T}\Big(-\text{tr}((\frac{\partial}{\partial\alpha_k}R)(\alpha^{\star},\beta^{\star},\delta^{\star})R(\alpha^{\star},\beta^{\star},\delta^{\star})^{-1})\\&+\text{tr}(S_TR(\alpha^{\star},\beta^{\star},\delta^{\star})^{-1} (\frac{\partial}{\partial\alpha_k}R)(\alpha^{\star},\beta^{\star},\delta^{\star})R(\alpha^{\star},\beta^{\star},\delta^{\star})^{-1})\Big)\nonumber\\&=\text{tr}\Big(-(\sqrt{T}(S_T-R(\alpha^{\star},\beta^{\star},\delta^{\star})))\\&\cdot(R(\alpha^{\star},\beta^{\star},\delta^{\star})^{-1}(\frac{\partial}{\partial\alpha_k}R)(\alpha^{\star},\beta^{\star},\delta^{\star})R(\alpha^{\star},\beta^{\star},\delta^{\star})^{-1})\Big)\nonumber\\&\stackrel{T\to\infty}\to\text{tr}\Big(-ZR(\alpha^{\star},\beta^{\star},\delta^{\star})^{-1/2}(\frac{\partial}{\partial\alpha_k}R)(\alpha^{\star},\beta^{\star},\delta^{\star})R(\alpha^{\star},\beta^{\star},\delta^{\star})^{-1/2}\Big),
\end{align*}} where $Z$ is a normal distributed random variable with mean $0_d0_d^\intercal$, covariances equal to 0 and variances equal to 1. $\sqrt{T}\max_W\{|\triangledown l(\alpha,\beta,\delta)|\}$ is bounded in probability by a similar argument, and $|(\alpha^{\star},\beta^{\star},\delta^{\star})-(\hat{\alpha}_{\text{SCE}},\hat{\beta}_{\text{SCE}},\hat{\delta}_{\text{SCE}})|$ converges to 0 by the first part of this theorem, which we have already proven. In total $\sqrt{T}R_T$ converges to 0 in probability and the assertion follows by Slutsky's theorem. 

We recognize one part of the Fisher information matrix in the above expression. The second part is obtained by taking covariances, such that the limit distribution of the gradient \\$\sqrt{T}(\frac{\partial}{\partial\alpha_k}l)(\alpha^{\star},\beta^{\star},\delta^{\star})$ ends up being a multivariate normal with mean 0 and covariance matrix equal to 4 times the Fisher information. In total, a repeated application of Slutsky's theorem and the continuous mapping theorem gives the multivariate normal with mean 0 and covariance matrix equal to the inverse of the Fisher information for the parameter vector, which was the assertion.
    
\end{proof}

\begin{proof}[Proof of Theorem 3.6]

Conditions for consistency and asymptotic normality are given in \citet[Theorem 2]{MaMa84-1sampleMLE}. One assumption of this theorem is Condition (A). The others are listed as Conditions (i)-(iv) in the original manuscript. We recognize the Fisher information matrix in Condition (iii), and since the mean of the standardized errors is equal to 0, there is no linear regression parameter to estimate and Condition (iv) is not necessary. To show that Condition (i) is fulfilled, we are first going to show that the spectral norm of $R$, $\|R\|_s$, defined by the largest singular value of $R$ (this the largest absolute value of the eigenvalues, since all of the following matrices are symmetric), is bounded in $d$. Since \begin{align}
    \|R\|_s\leq \delta\|\Gamma(\beta)^{-1}\|_s+\sum^K_{k=0}\alpha_k\|F_k\|_s\leq \|\Gamma(\beta)^{-1}\|_s+\sum^K_{k=0}\|F_k\|_s,
\end{align} it suffices to show that $\|\Gamma(\beta)^{-1}\|_s$ and $\|F_k\|_s$ are bounded in $d$. This follows by the Gershgorin circle theorem: all diagonal entries of the correlation matrices are equal to 1, so \begin{align}
    \|\Gamma(\beta)^{-1}\|_s\leq 1+\|\Gamma(\beta)^{-1}-I_d\|_s\leq 1+\max_i\sum_{j=1}^d|\Gamma(\beta)^{-1}_{i,j}|\leq V_{K+1}
\end{align} and \begin{align}
    \|F_k\|_s\leq 1+\|F_k-I_d\|_s\leq1+\max_i\sum_{j=1}^d|(F_k)_{i,j}|\leq V_k,
\end{align} for all $k$, where $V_0,\dots,V_{K+1}$ denote the bounds derived from Condition (C). These spectral norms are thus bounded. A similar argument can be made for the first and second partial derivatives of $R$, since $F_0=I_d$ (Condition C) and thus\begin{align}
    \|\frac{\partial}{\partial\alpha_k}R\|_s=\|-I_d+F_k\|_s,\quad \|\frac{\partial^2}{\partial\alpha_{k_1}\partial\alpha_{k_2}}R\|_s=\|\frac{\partial^2}{\partial\delta\partial\alpha_{k_2}}R\|_s=\|\frac{\partial^2}{\partial\delta^2}R\|_s=0
\end{align} for all $k,k_1,k_2$, and \begin{align}
    \|\frac{\partial}{\partial\beta_g}R\|_s=\delta\|(\frac{\partial}{\partial\beta_g}\Gamma(\beta))^{-1}\|_s,\quad \|\frac{\partial^2}{\partial\beta_{g_1}\partial\beta_{g_2}}R\|_s=\delta\|(\frac{\partial^2}{\partial\beta_{g_1}\partial\beta_{g_1}}\Gamma(\beta))^{-1}\|_s
\end{align} for all $\beta_g,\beta_{g_1},\beta_{g_2}$ and\begin{align}
    \|\frac{\partial^2}{\partial\alpha_{k}\partial\beta_{l}}R\|_s=0,\quad \|\frac{\partial^2}{\partial\delta\partial\beta_{l}}R\|_s\|=\|(\frac{\partial}{\partial\beta_{l}}\Gamma^{-1})(\beta)\|_s,\|\frac{\partial}{\partial\delta}R\|_s=\|\Gamma^{-1}(\beta)\|_s
\end{align} for all $k,l$. Note that the row-sum inequality still applies to the partial derivatives of $\Gamma^{-1}$: since the partial derivatives of all diagonal entries of $\Gamma^{-1}$ are 0, we can apply the Gershgorin circle theorem:\begin{align*}
    \|(\frac{\partial}{\partial\beta_g}\Gamma^{-1})(\beta)\|_s&\leq \max_i\sum^d_{j=1}|(\frac{\partial}{\partial\beta_g}\Gamma^{-1})(\beta)_{i,j}|,\\\|(\frac{\partial^2}{\partial\beta_{g_1}\partial\beta_{g_2}}\Gamma^{-1})(\beta)\|_s&\leq \max_i\sum^d_{j=1}|(\frac{\partial^2}{\partial\beta_{g_1}\partial\beta_{g_2}}\Gamma^{-1})(\beta)_{i,j}|.
\end{align*}

This implies Condition (i). Condition (ii) is true for $\Gamma(\beta)$ and its partial derivatives by assumption (Condition C). For the partial derivatives with respect to $\alpha_k$, it suffices to show that they are bounded from above by $O(\frac{1}{d})$. This is true: by Condition (C), there exists a $\tau>0$ such that for at least $p\in(0,1)$ percent of the off-diagonal row-sums of squares of $|F_k|$ are larger or equal to $\tau$ and thus\begin{align}
    \|\frac{\partial}{\partial\alpha_k}R\|_2^2=\|I_d-F_k\|_2^2\geq dp\tau\Leftrightarrow\frac{1}{\|\frac{\partial}{\partial\alpha_k}R\|_2^2}\leq\frac{1}{dp\tau},
\end{align} which proves the assertion.

\end{proof}

\begin{proof}[Proof of Theorem 3.8]

First, we note that the proof of Theorem 3.4 is exactly the same if the model is misspecified (Condition A3), except that we stated as a condition that the parameters converge (we needed to do that since the maximizer of $f$ is not necessarily unique in the misspecified setting). Thus, $\hat{R}_{SCE}$ still converges, but to $R(\alpha^\star,\beta^\star,\delta^\star)$ (follows from Condition B3), which is no longer equal to the true correlation matrix. Since $\beta^\star$ lies in the domain of $\Gamma$ (Condition C3), which is a bounded and open set (Condition B1), we can use the fact that $\Gamma$ is locally Lipschitz (follows from differentiability, Condition B2) to imply $\sqrt{T}$-convergence of $R(\hat{\alpha}_{\text{SCE}},\hat{\beta}_{\text{SCE}},\hat{\delta}_{\text{SCE}})$, since \begin{align*}
    &\textbf{E}[\sqrt{T}\|R(\hat{\alpha}_{\text{SCE}},\hat{\beta}_{\text{SCE}},\hat{\delta}_{\text{SCE}})-R(\alpha^\star,\beta^\star,\delta^\star)\|_2]\leq\\&\textbf{E}[\sqrt{T}\|R(\hat{\alpha}_{\text{SCE}},\hat{\beta}_{\text{SCE}},\hat{\delta}_{\text{SCE}})-R(\alpha^\star,\beta^\star,\delta^\star)\|_21_{\|(\hat{\alpha}_{\text{SCE}},\hat{\beta}_{\text{SCE}},\hat{\delta}_{\text{SCE}})-(\alpha^\star,\beta^\star,\delta^\star)\|_2\leq o_1}]+\\&\textbf{P}(\|(\hat{\alpha}_{\text{SCE}},\hat{\beta}_{\text{SCE}},\hat{\delta}_{\text{SCE}})-(\alpha^\star,\beta^\star,\delta^\star)\|_2> o_1)2d^2\max_{\alpha,\beta,\delta}\max_{i,j}R(\alpha,\beta,\delta)_{i,j}^2\leq\\& o_2\textbf{E}[\sqrt{T}\|(\hat{\alpha}_{\text{SCE}},\hat{\beta}_{\text{SCE}},\hat{\delta}_{\text{SCE}})-(\alpha^\star,\beta^\star,\delta^\star)\|_2]+\\&\textbf{P}(\|(\hat{\alpha}_{\text{SCE}},\hat{\beta}_{\text{SCE}},\hat{\delta}_{\text{SCE}})-(\alpha^\star,\beta^\star,\delta^\star)\|_2> o_1)2d^2\max_{\alpha,\beta,\delta}\max_{i,j}R(\alpha,\beta,\delta)_{i,j}^2\stackrel{T\to\infty}\to0,
\end{align*} where $o_1,o_2$ are two positive constants given by the Lipschitz-condition. This implies that $\gamma$ converges to a constant strictly larger than 0. The numerator is also well-defined, since $\sqrt{T}(\hat{R}_{\text{Pearson}}-R^\star)$ converges in $L^2$ (follows from Condition D3). It follows that $\pi,\rho$ are bounded in $T$ and $\gamma$ is bounded away from 0 in $T$. This implies the $O(\frac{1}{T})$ property of $\hat{\lambda}^\ast_{\text{WSCE}}$. For the other part of the assertion, we follow the proof of \cite{LeWo03-CovMatCombination}.

 Let $\hat{R}_1=R(\hat{\alpha}_{\text{SCE}},\hat{\beta}_{\text{SCE}},\hat{\delta}_{\text{SCE}})-R^\star$ and $\hat{R}_2=\hat{R}_{\text{Pearson}}-R^\star$. The expected squared error in terms of the Frobenius norm is given by \begin{align*}
    &\textbf{E}\left[\|(1-\lambda)R(\hat{\alpha}_{\text{SCE}},\hat{\beta}_{\text{SCE}},\hat{\delta}_{\text{SCE}})+\lambda\hat{R}_{\text{Pearson}}-R^\star\|_F^2\right]=\\\sum_{i\neq j}&\textbf{E}\left[\left((1-\lambda)(\hat{R}_1)_{i,j}+\lambda(\hat{R}_2)_{i,j}\right)^2\right]=\\\sum_{i\neq j}&\textbf{E}\left[\left((1-\lambda)(\hat{R}_1)_{i,j}+\lambda(\hat{R}_2)_{i,j}\right)\right]^2+\textbf{Var}\left[(1-\lambda)(\hat{R}_1)_{i,j}+\lambda(\hat{R}_2)_{i,j}\right]=\\\sum_{i\neq j}&a_{i,j}(1-\lambda)^2+b_{i,j}2(1-\lambda)\lambda+c_{i,j}\lambda^2,
\end{align*} with \begin{align*}
    Ta_{i,j}&=\textbf{Var}[\sqrt{T}(\hat{R}_1)_{i,j}]+\textbf{E}[\sqrt{T}(\hat{R}_1)_{i,j}]^2\\&=\textbf{Var}[\sqrt{T}(\hat{R}_{\text{SCE}}-R(\alpha^\star,\beta^\star,\delta^\star))_{i,j}]+\textbf{E}[\sqrt{T}(\hat{R}_1)_{i,j}]^2\\Tb_{i,j}&=\textbf{Cov}[\sqrt{T}(\hat{R}_1)_{i,j},\sqrt{T}(\hat{R}_2)_{i,j}]+\textbf{E}[\sqrt{T}(\hat{R}_1)_{i,j}]\textbf{E}[\sqrt{T}(\hat{R}_2)_{i,j}]\\&=\textbf{Cov}[\sqrt{T}(\hat{R}_{\text{SCE}})_{i,j},\sqrt{T}(\hat{R}_{\text{Pearson}})_{i,j}]+\textbf{E}[\sqrt{T}(\hat{R}_1)_{i,j}]\textbf{E}[\sqrt{T}(\hat{R}_2)_{i,j}]\\Tc_{i,j}&=\textbf{Var}[\sqrt{T}(\hat{R}_2)_{i,j}]+\textbf{E}[\sqrt{T}(\hat{R}_2)_{i,j}]^2\\&=\textbf{Var}[\sqrt{T}(\hat{R}_{\text{Pearson}})_{i,j}]+\textbf{E}[\sqrt{T}(\hat{R}_2)_{i,j}]^2.
\end{align*}

$\textbf{E}[\sqrt{T}(\hat{R}_2)_{i,j}]$ converges to 0, since $\hat{R}_2$ converges to a normal distribution with means 0 by Condition (C2) of Theorem 3.4. We can thus ignore this coefficient. The first order condition is\begin{align*}
    \sum_{i\neq j}-2a_{i,j}(1-\lambda)-2b_{i,j}\lambda+2b_{i,j}(1-\lambda)+2c_{i,j}\lambda=0,
\end{align*} with the solution \begin{align*}
    1-\hat{\lambda}^\ast_{\text{WSCE}}=\frac{-(\sum_{i\neq j}b_{i,j})+(\sum_{i\neq j}c_{i,j})}{(\sum_{i\neq j}a_{i,j})-2(\sum_{i\neq j}b_{i,j})+(\sum_{i\neq j}c_{i,j})}.
\end{align*}

Multiplying the numerator with $T$ and taking limits gives $\frac{1}{T}(\pi-\rho)+O(\frac{1}{T^2})$. The only expression in the denominator that does not go to 0 as $T\to\infty$ is \begin{align}
    \sum_{i\neq j}\textbf{E}[(\hat{R}_1)_{i,j}]^2=\sum_{i\neq j}\textbf{E}[(\hat{R}_{\text{SCE}})_{i,j}-R^\star_{i,j}]^2=\gamma.
\end{align}

This finishes the proof.

\end{proof}

\section*{Appendix C: details on the algorithm}

We only give details on the initialization step and on our estimation of the optimal mixing parameter of the WSCE, since the optimization step is straightforward. The only thing to point out about the optimization step is that our linear parameters are by nature constrained, since $\alpha_1,\dots,\alpha_K,\delta$ must lie in the $(K+1)$-dimensional simplex. We remove this constraint by mapping these parameters to $\mathbb{R}^{K+1}$ via a bijective softmax function, and adjust the derivative of the likelihood by the Jacobian of the derivative of this function. A similar thing can be done for $\beta$ if it is constrained as well. In our case, $\beta_D\in(0,1)$, so we used a sigmoid-function to map it to the real line. It should be noted that extremely large or small parameter estimates can cause numerical issues in the algorithm. In this case, we capped the estimates at an upper, respectively lower, threshold at which they did not cause problems.

\begin{algorithm}
\caption{Initialization Step}\label{alg:init_step}
\begin{algorithmic}
% \begin{itemize}
\item[]\textbf{Input:} matrices $F_1,\dots,F_K$, matrix function $\Gamma(\beta)^{-1}$, grid $(\beta^{(1)},\dots,\beta^{(N)})$, matrix estimator $\hat{R}$
% \newline
\item[]\textbf{Initialization:} 
\For{$(i,\beta^{(i)})\in\{(1,\beta^{(1)}),\dots,(N,\beta^{(N)})\}$}
\begin{align}\label{eq:quadprod_equation}
F^{(i)}&=\min_{\alpha_0,\dots,\alpha_K,\delta}\;(\alpha_0,\dots,\alpha_K,\delta)P(\alpha_0,\dots,\alpha_K,\delta)^\intercal - (\alpha_0,\dots,\alpha_K,\delta)h,\\&\text{s.t. }\alpha_0,\dots,\alpha_K,\delta\geq0\label{eq:quadprod_equation_constraint}
\end{align} where \begin{align*}
    &P_{i,j}=\begin{cases}
        \text{tr}(F_iF_j)&0\leq i,j\leq K,\\\text{tr}(F_i\Gamma(\beta^{(i)})^{-1})&0\leq i\leq K,j=K+1\text{ or }i=K+1,0\leq j\leq K,\\\text{tr}(\Gamma(\beta^{(i)})^{-1}\Gamma(\beta^{(i)})^{-1})&i=j=K+1,
    \end{cases}\\&h=(\text{tr}(F_1\hat{R}),\dots,\text{tr}(F_K\hat{R}),\text{tr}(\Gamma(\beta^{(i)})^{-1}\hat{R})).
\end{align*} 
\EndFor

\State Set $(\hat{\alpha}^{(0)}_0,\dots,\hat{\alpha}^{(0)}_K,\hat{\delta}^{(0)},\hat{\beta}^{(0)})=\text{argmin}_{(\alpha_0,\dots,\alpha_K,\beta,\delta)}\{F^{(1)},\dots,F^{(N)}\}$.

\While{$\exists n: (\hat{\alpha}^{(0)}_0,\dots,\hat{\alpha}^{(0)}_K,\hat{\delta}^{(0)})_n=0$}
    \State Set \begin{align*}
        q=\text{argmin}_k \sum_{i,j} |&((\lceil|F_0|\rceil,\dots,\lceil|F_K|\rceil,\lceil|\Gamma(\hat{\beta}^{(0)})^{-1}|\rceil)_k)_{(i,j)}\\-&((\lceil |F_0|\rceil,\dots,\lceil|F_K|\rceil,\lceil|\Gamma(\hat{\beta}^{(0)})^{-1}|\rceil)_n)_{(i,j)}|.
    \end{align*} 
    \State Add the constraint $(\alpha,\delta)_k\geq\max\{\frac{(\hat{\alpha},\hat{\delta})_q^{(0)}}{K+2},\exp(-15)\}$ to Restriction \eqref{eq:quadprod_equation_constraint}.
    \State Re-initialize. 
\hspace{1cm} 

\EndWhile
\item[]\textbf{Output:} \begin{align*}
    (\hat{\alpha}^{(0)},\hat{\beta}^{(0)},\hat{\delta}^{(0)}).
\end{align*}

\end{algorithmic} 
\end{algorithm}

A detailed description of the initialization step used to compute the SCE is given in pseudo-code form in Algorithm 1. The quadratic optimization problem \eqref{eq:quadprod_equation} is solved by a method presented in \citet{GoId83-quadratic_optimization02, GoId06-quadratic_optimization01}, that is implemented in the quadprog package \citep{BeWe19-quadprog}. The method assumes that the matrix $P$ is positive definite. This is in fact the case, if the conditions of the first theorem hold.

The definition of Problem \eqref{eq:quadprod_equation} follows from the fact that \begin{align*}
    \|R(\alpha,\beta,\delta)-\hat{R}\|_F^2&=\delta\Gamma(\beta)^{-1}+\|\sum^{K}_{k=0}\alpha_k F_k-\hat{R}\|_F^2\\&\propto(\alpha_0,\dots,\alpha_K,\delta)P(\alpha_0,\dots,\alpha_K,\delta)^\intercal-(\alpha_0,\dots,\alpha_K,\delta)h.
\end{align*}

The idea behind the restrictions added in the case that the solution lies on the edge of the parameter space is that in practice, we have observed that sometimes $\alpha_k=0$ and $\alpha_k=1-\delta-\sum_{m\in[K]\backslash\{k,n\}}\alpha_m$ if the matrices $F_k$ and $F_n$ are very similar. Thus, we force the component that was set to 0 to be strictly larger than a proportion of the component which has the most similar $"$support$"$ with respect to the $L^1$-norm, where by $"$support$"$ we mean the non-negative entries of each matrix (for correlation matrices, this corresponds to taking the absolute value and rounding). If the coefficient of the matrix closest to it is also 0, we choose a constant (exp(-15)) as the constraint of the parameter. The division by $K+2$ assures that the set defined by Equation \eqref{eq:quadprod_equation_constraint} is non-empty, since the sum of all added coefficients is always smaller or equal to 1.

\paragraph*{\textbf{Our estimators of $\pi,\rho,\gamma$}}

If we do not use bootstrap, we use \begin{align}\label{eq:pi_estim}
    \hat{\pi}=\sum^{d}_{i,j}\frac{(1-(\hat{R}_{\text{Pearson}})_{i,j}^2)^2}{T-1},
\end{align}which was e.g., used in \citet{Ha53-standard_corr_var_estim}, and\begin{align}\label{eq:gamma_estim}
    \hat{\gamma}=\sum^d_{i,j}((\hat{R}_{SCE})_{i,j}-(\hat{R}_{\text{Pearson}})_{i,j})^2,
\end{align} where Equation \eqref{eq:gamma_estim} was used in the original paper of \citet{LeWo03-CovMatCombination}. $\rho_U$ is then estimated using the same variance estimate as Equation \eqref{eq:pi_estim} and an estimate of the asymptotic variance of $\hat{R}_{\text{SCE}}$, obtained by the delta method.

If we do use bootstrap, we sample 100 different standardized errors $\varepsilon^{(1)},\dots,\varepsilon^{(100)}\stackrel{i.i.d}\sim\text{MVN}_d(0_d,\hat{R}_{\text{Pearson}})$ and calculate $(\hat{\pi},\hat{\rho},\hat{\gamma})$ via empirical estimators.

\section*{Appendix D: alternative model justifications}

In the main document we showed that we could write the correlation matrix $R$ of our model as a convex combination of known correlation matrices, 

\begin{equation} \label{AppendixD: model}
R(\alpha,\beta,\delta) = \Phi(\alpha) + \delta\cdot \Gamma(\beta)^{-1}, \qquad
\text{where} \quad
\Phi(\alpha) = \sum_{k=0}^K \alpha_k F_k, \quad
%\quad \alpha_k,\beta_\ell>0,
\end{equation}
with the constraints \begin{equation}
    \alpha_k,\delta>0,\quad\delta+\sum^K_{k=0}\alpha_k=1.
\label{AppendixD: eq:general_simplex_restriction}\end{equation}

We justified this by presuming that the standardized errors of our data, $\varepsilon$, are themselves a weighted sum of standardized, independent effects. However, there exist alternative justifications for using our approach. The fact that multiple justifications exist improves the construct-validity of our model, since all of these distinct and very common justifications are arguments for its use. One justification is to say that each matrix $F_1,\dots,F_K$ belongs to a multilevel factor model, and that $R$ is an average over all of those models and the CAR model. Another is to write our model as a linear mixed-effects model. We are going to elaborate on these justifications here.

\paragraph*{\textbf{Model-averaged multilevel factor models}}

Suppose that we have not one, but a variety of competing models $\mathcal{M}_0,\dots,\mathcal{M}_{K+1}$, each corresponding to a Gaussian distribution with correlation matrix $F_0,\dots,F_K,\Gamma(\beta)^{-1}$, respectively. Then, for any fixed value of $\beta$, the model-averaged estimator \citep{AnBu04-model_averaging} is \begin{align}
    \hat{R}(\mathcal{M}_0,\dots,\mathcal{M}_{K+1})=\omega_{K+1}\Gamma(\beta)^{-1}+\sum^{K}_{k=0}\omega_k F_k.
\end{align}

We are going to show that we obtain exactly the same estimator when using Equation \eqref{AppendixD: model}, and we are going to give the models $\mathcal{M}_0,\dots,\mathcal{M}_{K+1}$ by which we arrive at this suggestion. 

First, we are going to define the models of the non-spatial components, $\mathcal{M}_0,\dots,\mathcal{M}_K$, as a special case of multilevel models, where the group-means are known. In the main document we set $F_0=I_d$ and  wrote each of the matrices with the capital letter $F$ as $F_k=f_k^\intercal f_k$, where $f_k$ are binary matrices that denote the membership in different clusters. Multilevel factors models are another way of arriving at the design of $F_k$. 

The covariance matrix in multilevel factor models is a sum of two parts: The between-group covariance matrix, which is a result of different groups having different means, and the within-group covariance matrix. However, the mean-structure of our model is already known and does not depend on the groups, so let us assume that the covariance matrix of $F_k$, $\tilde{F}_k$ can be modelled as the covariance matrix of a multilevel model with a between-group variance equal to 0 and a within-group covariance matrix. If we further assume that the covariance coefficient between two members of the same cluster is always the same (this would account to a 1-factor model with a factor of ones, multiplied by a positive constant $c_k$), then\begin{align}
    (\tilde{F}_k)_{i,j}=\begin{cases}
        s^2+c_k&i=j,\\c_k&i\neq j,\text{ i and j belong to the same cluster},\\0&\text{else,}
    \end{cases}
\end{align} where $s^2$ is the variance parameter. Substituting $\alpha_k'=\frac{c_k}{s^2+c_k}$ gives a correlation matrix of \begin{align}
    (1-\alpha_k')F_0+\alpha_k' F_k=\begin{cases}1&i=j,\\
        \alpha_k'&i\neq j,\text{ i and j belong to the same cluster,}\\0&\text{else.}
    \end{cases}
\end{align}

This observation already implies our assertion: averaging all of these $K$ correlation matrices and the correlation matrix of the CAR model, $\Gamma(\beta)^{-1}$, with a vector $\omega_0,\dots,\omega_K,\omega_{K+1}>0$ such that $\sum^{K+1}_{k=0}\omega_k=1$, gives \begin{align}
    R(\alpha,\beta,\delta)=\omega_{K+1}\Gamma(\beta)^{-1}+\sum^{K}_{k=0}\omega_k\left((1-\alpha_k')F_0+\alpha_k'F_k\right)=\delta\cdot\Gamma(\beta)^{-1}+\sum^K_{k=0}\alpha_kF_k,
\end{align} if we substitute $\alpha_k=\alpha_k'\omega_k$ for $1\leq k\leq K,$ $\delta=\omega_{K+1}$, $\alpha_0=1-\delta-\sum^K_{k=1}\alpha_k$. Interestingly, this shape cannot be obtained when using sums of covariance matrices instead of correlation matrices, as is e.g., done in linear mixed-effects models. However, if we separate the correlation-matrix estimation from the variance estimation, one could also arrive at our model when using the framework of these models.

\paragraph*{\textbf{Linear mixed-effects models}} Linear mixed-effects models can, in their broadest form, be written as \begin{align}\label{eq: mixed models}
    Y=X\beta' + Zu+\varepsilon',
\end{align} where $\beta'$ is an unknown constant, $u$ follows a multivariate normal distribution with a mean vector of zeroes, $X,Z$ are known design matrices, and $\varepsilon'$ is i.i.d Gaussian noise with mean 0 and variance $\sigma^2_E$. Let us split $u$ into $u_A,\dots,u_D$ and $Z$ into $Z_A,\dots,Z_D$, such that Equation \eqref{eq: mixed models} becomes \begin{align}\label{eq:mixed models with ZA,...,ZD}
    Y=X\beta' + Z_Au_A+Z_Bu_B+Z_Cu_C+Z_Du_D+\varepsilon'.
\end{align}

Now let's assume that all entries of $u_A,\dots,u_D$ are independent with variances $\sigma_A^2,\dots,\sigma_D^2$ and let $Z_A=f_A,Z_B=f_B,Z_C=1_d,Z_D=\sqrt{\Gamma(\beta_D)^{-1}}$, where $\sqrt{\Gamma(\beta_D)^{-1}}$ denotes the Cholesky decomposition. Then $\Sigma$, the covariance matrix corresponding to Equation \eqref{eq:mixed models with ZA,...,ZD} is given by\begin{align}
    \Sigma=\sigma_A^2 f_Af_A^\intercal+\sigma_B^2f_Bf_B^\intercal+\sigma_C^21_d1_d^\intercal+\sigma^2_EI_d.
\end{align} Substituting \begin{align}
    (\alpha_A,\dots,\delta_D,\alpha_E)=(\sigma_A^2,\dots,\sigma^2_E)\frac{1}{\sigma_A^2+\dots+\sigma^2_E},\\F_A=f_Af_A^\intercal,\quad F_B=f_Bf_B^\intercal,\quad F_C=1_d1_d^\intercal,\quad\Gamma(\beta_D)^{-1}=f_Df_D^\intercal,
\end{align} gives the correlation matrix\begin{align}
    R(\alpha,\beta,\delta)=\alpha_AF_A+\alpha_BF_B+\alpha_C F_C+\delta_D\Gamma(\beta_D)^{-1}+\alpha_E I_d,
\end{align} which corresponds to our model introduced in the main document. Thus, for any fixed parameter $\beta_D$, the correlation structure of this model can be expressed as the correlation structure of a linear mixed-effects model. Hadamard products such as $F_{A,B}=F_A\odot F_B$ also get a nice interpretation in this setting: let $f_{A,B}=f_A\newmoon f_B$, where $\newmoon$ denotes the facesplitting product, a product between the covariates $f_A$ and $f_B$. This gives a new covariate, which will result in $F_{A,B}$ as a correlation matrix.

\section*{Appendix E : simulation settings details}
\paragraph*{\textbf{Randomly simulated correlation structures}}

%{{\textbf{Setting}}}\\
We set $(d,T)=(200,10)$. The matrices $F_A$ and $F_B$, which define the correlation matrices of the common colonizer and the regional effect, respectively, were simulated via the multinomial distribution. We simulated the random matrices $\tilde{f}_A\in\{0,1\}^{d\times 3},\tilde{f}_B\in\{0,1\}^{d\times 10}$, such that \begin{align}
    F_A&=\tilde{f}_A \tilde{f}_A^\intercal,\quad \tilde{f}_A\sim \prod^d_{i=1}\text{multinom}(1,\frac{1}{3},\frac{1}{3},\frac{1}{3}),\\
    F_B&=\tilde{f}_B \tilde{f}_B^\intercal,\quad \tilde{f}_B\sim \prod^d_{i=1}\text{multinom}(1,\frac{1}{10},\dots,\frac{1}{10}),\\
    F_C &= 1_d 1_d^\intercal, \quad 1_d = \{1\}^{d \times 1}.
\end{align} 

The adjacency matrix $M$ was simulated such that it follows the law of an adjacency matrix of an Erdős-Rényi random graph \citep{ErRe1959randomgraphs}. The connection probability was chosen to be $\log(d)/d$.

We set $\alpha=(\alpha_A,\alpha_B,\alpha_C)=(0.05,0.09,0.11)$ (coefficients of $"$common colonizer$"$, $"$same region$"$, $"$global effect$"$). For the contiguity effect (in the following also referred to as $"$neighborhood effect$"$), we chose $(\delta_D,\beta_D)=(0.74,0.982)$, such that the largest neighborhood effect is roughly equal to 0.26. The values were chosen such that they closely match the parameter values calculated in \cite{FoRa14-corlinearreg}. The mean and variance parameters were set to be $0$ and $1$, respectively. 

\paragraph*{\textbf{Using the correlation structures from the data}}

$\mu,\sigma$, are presumed to be known and set to $\mu\equiv0,\sigma\equiv1$. We chose $(\alpha_A,\alpha_B,\alpha_C,\delta_D,\beta_D)=(0.11,0.05,0.09,0.01,0.35)$ to closely resemble the values found in \cite{FoRa14-corlinearreg} and repeated 40 independent simulations, evaluating the same estimators that were presented in the previous section.  

\paragraph*{\textbf{Performance when missing values are present}}

We repeat the same simulations, but with the missing values present in the data. The missing value structure $O_t\subseteq\{1,\dots,d\}$ was taken from the original dataset and it was presumed that\begin{align}
    Y_t\sim\text{MVN}_d(\mu_{O_t},(\text{diag}(\sigma_t)\Sigma(\alpha,\beta,\delta)\text{diag}(\sigma_t))_{O_t,O_t})
\end{align} independently for every $t=1,\dots,T$. The presence of missing values is not a problem for the SCE, since we can still maximize over the product of the approximate likelihood values\begin{align}\label{eq:likeli_missingvalues}
    L(\alpha,\beta,\delta)=\prod^T_{t=1}\text{MVN}_d(Y_{O_t},\hat{\mu}_{O_t},(\text{diag}(\hat{\sigma}_t)R(\alpha,\beta,\delta)\text{diag}(\hat{\sigma}_t))_{O_t,O_t}).
\end{align} Notice that this corresponds to $"$marginalizing out$"$ the missing values of the dataset. This is appropriate if the values of the dataset are missing at random everywhere \citep{Ru76-missing_at_random,Sh_et_al13-missingdata}. We make the case for this in the main document. Constructing the IVE is not a problem either, since we can just set it to minimize the distance to the $"$Pearson correlation matrix$"$ that was obtained by individually estimating the correlation coefficient of each country pair, for all data pairs that were available. This $"$Pearson correlation matrix$"$ is useful for constructing the IVE, but not in of itself. This is the case because it is not positive definite, a consequence of estimating each correlation coefficient individually. The results, for which 40 simulations were repeated independently, using the missing value structure from the dataset, do not change much.

\begin{figure}
\centering
\includegraphics[width=243.6pt]{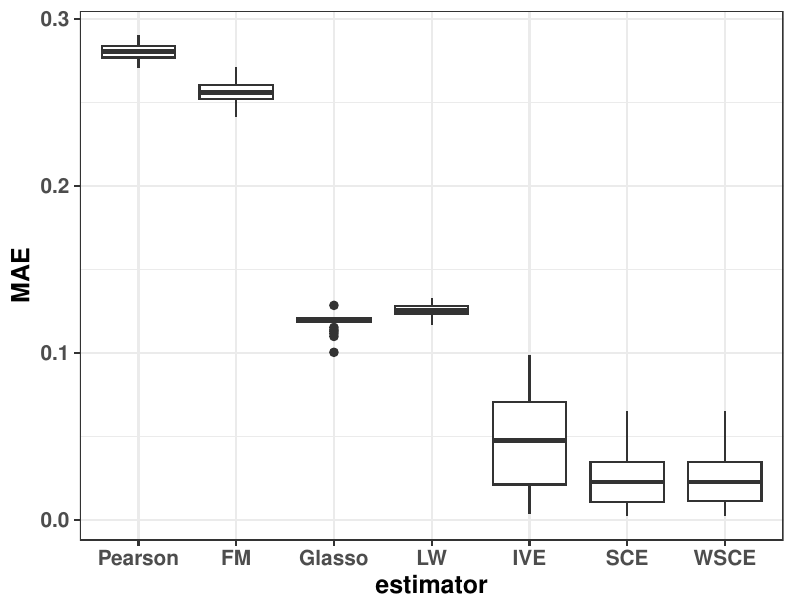}
\caption{{Boxplots of the mean absolute error (MAE) for the Pearson correlation matrix (Pearson) the estimator that uses factor models (FM), the glasso estimator (Glasso) the Ledoit-Wolf estimator (LW), and the IVE, SCE, and WSCE. Estimators were evaluated in the fully simulated setting (FSS) and the setting of the TFR dataset (TFR). The presence of missing values has a minor impact on all estimators. The WSCE was also always equal to the SCE in all simulations, which may be a result of the increasing variance of the Pearson correlation estimator.}}\label{fig: sim_03_error_measures}
\end{figure}

The results are shown in Figure \ref{fig: sim_03_error_measures}. All of the previous state-of-the-art estimators require all data to be present, so we imputed the missing data. The standardized errors of the data were imputed using PCA, with a method that was implemented in \citet{JoHu16-missingvalueimputation} and the number of principal components set to the number of model parameters.

\paragraph*{\textbf{Missing values and model misspecification}}

We set\begin{align*}
    R_{miss}&=\xi R(\alpha,\beta,\delta)+(1-\xi)\tilde{R},
    \\\tilde{R}&=0.01 I_d+0.99F_{miss},
    \\(F_{miss})_{i,j}&=(\tilde{f}_A\tilde{f}_A^\intercal)_{i,j},\quad i,j=1,\dots,195,
\end{align*}meaning that we add up to $99\%$ of an unknown covariate,$\tilde{f}_A$, to the model, which was simulated in the setting defined in the beginning of this section. For both of these experiments, we repeat 10 independent simulations for $\xi$ varying between $\xi=0,0.1,\dots,0.9,1$.

As before, we computed the SCE by maximizing the likelihood, for which the missing values were marginalized out via Equation \eqref{eq:likeli_missingvalues}, and computed the Pearson correlaton matrix on the imputed dataset. Computing $\hat{\lambda}_{\text{WSCE}}$ is less straightforward: due to the data imputation, we can not use the standard estimator of the asymptotic variance of the Pearson correlation matrix. Indeed, we found that computing this estimate on the imputed dataset produced very inaccurate estimates. Instead, we used parametric bootstrap to calculate the optimal shrinkage parameter $\hat{\lambda}_{\text{WSCE}}^*$. The IVE was used directly to initialize each sample to shorten computation time, as suggested in \cite{Ch23-bootstrap_init}.

\bibliographystyle{chicago}
\bibliography{biblio.bib}

% \bibliographystyle{chicago}
% \bibliography{biblio.bib}